\documentclass[twocolumn]{aastex63}

\newcommand{\mytilde}{\raise.19ex\hbox{$\scriptstyle\sim$}}

\renewcommand{\d}{{\mathinner{\mathrm{d}}}}

\usepackage[utf8]{inputenc}
\usepackage{CJK}
\usepackage{amsmath}

\received{}
\revised{}
\accepted{}
\shorttitle{Deep Learning for Weak-lensing Mass Reconstruction}
\shortauthors{Hong et al.}
\graphicspath{{./}{figures/}}

\begin{document}

\title{Weak-lensing Mass Reconstruction of Galaxy Clusters with Convolutional Neural Network}

\correspondingauthor{M. James Jee}
\email{mkjee@yonsei.ac.kr}

\author[0000-0003-4923-8485]{Sungwook E. Hong \begin{CJK*}{UTF8}{mj}(홍성욱)\end{CJK*}}
\affiliation{Natural Science Research Institute, University of Seoul,
163 Seoulsiripdaero, Dongdaemun-gu, Seoul 02504, Republic of Korea}
\affiliation{Korea Astronomy and Space Science Institute,
776 Daedeok-daero, Yuseong-gu, Daejeon 34055, Republic of Korea}

\author{Sangnam Park}
\affiliation{Natural Science Research Institute, University of Seoul,
163 Seoulsiripdaero, Dongdaemun-gu, Seoul 02504, Republic of Korea}

\author[0000-0002-5751-3697]{M. James Jee}
\affiliation{Department of Astronomy, Yonsei University,
50 Yonsei-ro, Seodaemun-gu, Seoul 03722, Republic of Korea}
\affiliation{Department of Physics, University of California, Davis, One Shields Avenue, Davis, CA 95616, USA}

\author[0000-0001-6764-3236]{Dongsu Bak}
\affiliation{Natural Science Research Institute, University of Seoul,
163 Seoulsiripdaero, Dongdaemun-gu, Seoul 02504, Republic of Korea}
\affiliation{Department of Physics, University of Seoul,
163 Seoulsiripdaero, Dongdaemun-gu, Seoul 02504, Republic of Korea}

\author{Sangjun Cha}
\affiliation{Department of Astronomy, Yonsei University, 
50 Yonsei-ro, Seodaemun-gu, Seoul 03722, Republic of Korea}



\begin{abstract}
We introduce a novel method for reconstructing the projected  matter distributions of galaxy clusters with weak-lensing (WL) data based on convolutional neural network (CNN). 
Training datasets are generated with ray-tracing through cosmological simulations. We control the noise level of the galaxy shear catalog such that it mimics the typical properties of the existing ground-based WL observations of galaxy clusters.
We find that the mass reconstruction by our multi-layered CNN with the architecture of alternating convolution and trans-convolution filters significantly outperforms the traditional reconstruction methods.  The CNN method provides better pixel-to-pixel correlations with the truth, restores more accurate positions of the mass peaks, and more efficiently suppresses artifacts near the field edges.
In addition, the CNN mass reconstruction lifts the mass-sheet degeneracy when applied to our projected cluster mass
estimation from sufficiently large fields.
This implies that this CNN algorithm can be used to measure cluster masses in a model-independent way for future wide-field WL surveys.  

\end{abstract}



\section{Introduction}\label{sec:intro}
Weak-lensing (WL) is now firmly established as the most direct method to measure the mass of an astrophysical object ranging from a galaxy (galaxy-galaxy lensing) to the cosmological large scale structure (cosmic shear). 
Many on-going and future WL surveys happening on massive scales reflect the elevated level of interest and confidence in this technique \citep[e.g.,][]{hikage2019,troxel2018,ivezic2019,euclid2011,wfirst2015}. 
Without question, in order to maximize the scientific return from these huge data volume, our highest priority is to understand and control systematics.
A number of issues on WL systematics have been identified, including shear calibration, photometric redshift degeneracy, model bias, mass-sheet degeneracy, astrophysical processes, and so on \citep[e.g.][]{GFS88,Seitz:1995dq,Squires:1995gs,high2007,jarvis2008,meyers2015, mandelbaum2015}.

In this study, we focus on systematics arising in galaxy cluster mass reconstruction from WL source catalogs. Although galaxy cluster mass reconstruction is one of the earliest WL application and demonstration of its power, the main utility of the two-dimensional mass reconstruction has been rather qualitative investigation of the relative mass distribution of the target field. Very few studies employed the mass reconstruction for quantitative analysis (e.g., derivation of galaxy cluster masses). This is because the current mass reconstruction algorithms suffer from various artifacts.
For example, the so-called mass-sheet degeneracy (invariant of the observable shear under a certain linear rescaling of the mass) is one of the major obstacles that prevent us from interpreting the result in absolute terms.
Severe nonlinearity arising from the transformation of the reduced shear to the convergence is also a crucial contributing factor. Other critical issues include finite-field effect, ill-posed mathematical inversion, smoothing artifact, field edge systematics, and so on \citep[e.g.,][]{Bartelmann:1995yq,Seitz:1995dq}

The most popular method to estimate the cluster mass so far has been to fit an analytic profile to the observed shear. This assumes that the cluster mass distribution is spherically symmetric and follows a particular halo model favored by numerical simulations such as an Navarro-Frenk-White \citep{navarro1996} profile. Although this provides a method to overcome the aforementioned drawbacks of the mass reconstruction, the obvious weakness is that individual galaxy clusters do not exactly follow the analytical description, much less are consistent with the assumption of  spherical symmetry. 

In this paper, we introduce a novel method for mass reconstruction based on convolutional neural network (CNN). 
CNN is a branch of deep learning, which has been considered to be a promising tool in many fields of astronomy in recent years such as photometric redshift \citep[e.g.,][]{schaefer2018}, strong-lensing finding \citep[e.g.,][]{pasquet2019}, image deconvolution \citep[e.g.,][]{flamary2017}, star-galaxy separation \citep[e.g.,][]{kim2017}, morphological classification \citep[e.g.,][]{mittal2019}, etc.
The current study is the first endeavor to apply CNN to WL mass reconstruction of galaxy clusters. Since there is a rapid growth in data size and complexity from future WL surveys, the approach introduced here will find many useful applications if our CNN algorithm can significantly reduce aforementioned systematics found in the traditional mass reconstruction.

This paper is organized as follows.
In \textsection\ref{sec:method}, we describe the basic theory, CNN architecture, and training data sets. The performance of our CNN mass reconstruction
is presented in \textsection\ref{sec:results} and discussed in \textsection\ref{sec:discussion} before
we conclude in \textsection\ref{sec:conclusion}. Throughout the paper, we assume a flat $\Lambda$CDM cosmology with $H_0=70$ km s$^{-1}$ Mpc$^{-1}$, $\Omega_\Lambda=0.7$, and $\Omega_{\rm m}=0.3$.


\section{Methods} \label{sec:method}
\subsection{Basic Weak-lensing Theory}
The basic WL theory is briefly reviewed here to make our method description self-contained. We refer readers to other excellent review papers for further details \citep[e.g.,][]{mellier1999,bartelmann2001,hoekstra2013}.
WL formalism is valid in the regime where the source galaxy is much smaller
than the characteristic scale of the gravitational potential variation. In this regime, the transformation matrix $\mathbf{A}$ relating the source plane position $\mathbf{x}$ to the image plane position $\mathbf{x}^{\prime}$ via $\mathbf{x^{\prime}}=\mathbf{A} \mathbf{x}$ is described by:
\begin{equation}
    \mathbf{A}= (1-\kappa)
    \begin{pmatrix}
    1-g_1  & -g_2 \\
    -g_2 & 1 + g_1
    \end{pmatrix},  \label{eqn:A}
\end{equation}
where $g_{1(2)}$ is the first (second) component of the reduced shear $g=(g_1^2+g_2^2)^{1/2}$ and $\kappa$ is the convergence.
The reduced shear $g$ is related to shear $\gamma$ and convergence $\kappa$ via 
\begin{equation}
g=\gamma/(1-\kappa).
\end{equation}
The convergence $\kappa$ is the unitless surface mass density:
\begin{equation}
\kappa=\frac{\Sigma}{\Sigma_c},    
\end{equation}
\noindent
where $\Sigma_c$ is the critical surface mass density:
\begin{equation}
\Sigma_c=\frac{c^2 D_{s}}{4\pi G D_l D_{ls}}. \label{eqn:sigma_c}
\end{equation}
In equation~(\ref{eqn:sigma_c}), $c$ is the speed of light, $G$ is the gravitational constant, $D_l$ is the angular diameter distance to the cluster (lens),
$D_{ls}$ is the angular diameter distance from lens to source, and $D_s$ is the angular diameter distance to source.

The transformation matrix $\mathbf{A}$ in equation~(\ref{eqn:A}) converts a circle into an ellipse. There are multiple ways to define the ellipticity of the resulting ellipse, which has been a source of confusion.
If we let its semi-major and -minor axes be $a$ and $b$, respectively, one can show that the reduced shear $g$ in equation~(\ref{eqn:A}) becomes
\begin{equation}
g=\frac{a-b}{a+b}. \label{eqn:e_definition}
\end{equation}
Therefore, it is convenient to use equation~(\ref{eqn:e_definition}) to define the ellipticity in WL, which we also adopt in this paper.
Since $g$ alone cannot express the orientation of the ellipse, the WL community often uses the complex notation: 
\begin{equation}
    \mathbf{g}=g_1 + \mathbf{i} g_2, \label{eqn:complex}
\end{equation}
which provides both magnitude $g=(g_1^2+g_2^2)^{1/2}$ and orientation $\phi=0.5 \tan^{-1}(g_2/g_1)$ of the elongation.

Under the assumption that we can assign a unique ellipticity to every galaxy, the same complex notation (equation~\ref{eqn:complex}) can also be used to express its  intrinsic ellipticity $\mathbf{e}=e_1 + \mathbf{i} e_2$ prior to WL distortion. Then, the transformation of the intrinsic ellipticity $\mathbf{e}$ to the lensed (distorted) ellipticity $\boldsymbol{\epsilon}$ by the reduced shear $\mathbf{g}$ is given by:
\begin{equation}
    \boldsymbol{\epsilon} = \frac{ \mathbf{e} + \mathbf{g}} { 1 + \mathbf{g}^* \mathbf{e}} \label{eqn:e_transform} ~\mbox{for}~ |\mathbf{g}|<1
\end{equation}
and
\begin{equation}
    \boldsymbol{\epsilon} = \frac{1 + \mathbf{g} \mathbf{e}^* } { \mathbf{e}^* + \mathbf{g}^*} \label{eqn:e_transform2} ~\mbox{for}~ |\mathbf{g}|>1,
\end{equation}
\noindent
where the symbol $^*$ denotes the complex conjugate.

Inspection of equation (\ref{eqn:e_transform}) reveals that in general each galaxy's lensed ellipticity $\boldsymbol{\epsilon}$ is only slightly different from its intrinsic ellipticity $\mathbf{e}$ in the WL regime where $\mathbf{g}$ is small. When we disregard measurement systematics and assume that the ellipticity distribution of the source population is isotropic, one can show that the unbiased estimator for $\mathbf{g}$ is $\left < \boldsymbol{\epsilon} \right >$.

\subsection{Conventional Mass Reconstruction and its Limitation}
\label{sec:conventional_MR}
The mathematical relation between $\boldsymbol{\gamma}$ shear and convergence $\kappa$ at the position $\mathbf{x}$ is:
\begin{equation}
    \boldsymbol{\gamma} (\mathbf{x}) = \frac{1}{\pi} \int \mathbf{D}(\mathbf{x}-\mathbf{x}^{\prime}) \kappa (\mathbf{x}^{\prime}) {\rm d} \mathbf{x}^{\prime}, \label{eqn:kappa2shear}
\end{equation}
where the kernel $\mathbf{D}$ is:
\begin{equation}
    \mathbf{D}=- \frac{1}{ (x_1- \mathbf{i} x_2)^2 }.
\end{equation}
The well-known \citet[][KS93]{kaiser1993} mass reconstruction is based on the straightforward inversion of equation~(\ref{eqn:shear2kappa}):
\begin{equation}
    \kappa (\mathbf{x}) = \frac{1}{\pi} \int \mathbf{D}^{*}(\mathbf{x}-\mathbf{x}^{\prime}) \boldsymbol{\gamma} (\mathbf{x}^{\prime}) {\rm d} \mathbf{x}^{\prime}. \label{eqn:shear2kappa}
\end{equation}
KS93 evaluate this convolution in Fourier space while \citet{fischer1997} develop an inversion method in real space. Alternatively, some authors propose to reconstruct the convergence field using equation~(\ref{eqn:kappa2shear}) through the maximum likelihood method \citep[e.g.,][]{Seitz:1998mg,bradac2004,jee2007}.

Inspection of equations~(\ref{eqn:kappa2shear}) and (\ref{eqn:shear2kappa}) shows that several artifacts may be introduced from the KS93 mass reconstruction. First, the evaluation of the convolution suffers from the so-called finite-inversion problem because in principle $\kappa$ requires the information of the shear $\boldsymbol{\gamma}$ over an infinite area. Second, equation~(\ref{eqn:shear2kappa}) uses $\boldsymbol{\gamma}$ for its input whereas the directly attainable information from averaging over many galaxy shapes is only $\mathbf{g}$. Third, the solution is not unique as the same equation holds under the transformation: $\kappa \rightarrow \lambda \kappa + (1 -\lambda)$, where $\lambda$ is arbitrary. This ambiguity is often termed the ``mass-sheet degeneracy'' because, although mathematically somewhat misleading, one can view the transformation as an addition of a thin sheet of mass when $\lambda \approx 1$.
Fourth, in the central region of massive clusters where the WL assumption no longer holds, the above equations lose their validity. Fifth, to suppress noise amplification in the inversion, we can only obtain a smoothed convergence field, which gives a biased mass estimate even in the ideal situation where all other issues are carefully accounted for.


\subsection{Mass Reconstruction with Convolutional Neural Network}\label{sec:method_cnn}

\subsubsection{Generation of Training Dataset}
\label{sec:method_sim}
Generation of our training dataset starts from the convergence ($\kappa$) maps created from cosmological simulations via ray tracing. We utilize the publicly available data MassiveNuS \citep{liu2018}. The simulation was originally designed to investigate the impact of massive neutrinos on the large-scale structure. For the current investigation, we
chose to retrieve\footnote{\url{http://columbialensing.org}} the dataset corresponding to the  $\sum m_{\nu}=0.177$~eV, $\Omega_{\rm m}=0.2485$, and  $A_s=2.0644\times10^{-9}$ setting.
However, we emphasize that the details in the simulation parameters are not important within the scope of the current study because, as we shall demonstrate later, our CNN algorithm is designed to learn the rule that maps the reduced shear field to the convergence field  according to general relativity and thus is independent of the above cosmological parameters.

The original dataset consists of a total of 50,000 convergence images at five ($z=0.5$, 1.0, 1.5, 2.0, and 2.5) different source redshifts (10,000 convergence fields per source redshift) each simulating an area of $3.5^\circ \times 3.5^\circ$ with a pixel resolution of $0\farcm4$, which matches the field of view (FOV) of the Vera C. Rubin Observatory \citep{ivezic2019}. The large convergence pixel makes the maximum convergence value never exceed unity.
For the same field, a higher source redshift convergence image is richer in substructure as more line-of-sight (LOS) structure is included and also the lensing efficiency becomes higher. We use the dataset created for the source redshift of 1.5. We verified that training with all five source redshift data does not improve the result.

We identified clusters by running SExtractor \citep{bertin1996} on the convergence field image and cropped a $32\arcmin \times 32\arcmin$ region approximately centered on each cluster. We randomly generated positions of 25,000 sources within this subfield. The distribution matches the typical source density of $\mytilde25$ per sq. arcmin in our previous Subaru WL studies \citep[e.g.,][]{finner2017, kim2019, yoon2020}.
Shears $\boldsymbol{\gamma}$ at the position $\mathbf{x}$ were computed using equation~(\ref{eqn:kappa2shear}).
This shear $\boldsymbol{\gamma}$ is then converted to
the reduced shear $\mathbf{g}$ through $\mathbf{g}=\boldsymbol{\gamma}/(1-\kappa)$. 
Here we do not consider dispersions in both lens and source redshift and assume that all lensing masses and sources are confined to $z_l=0.5$ and $z_s=1.5$, respectively. 
Also, as explained above, no convergence pixel exceeds unity, and thus only equation~(\ref{eqn:e_transform}) is needed for the ellipticity transformation.

We set the intrinsic shape noise per component to $\sigma_e=0.24$, which is approximately the empirical value from \emph{Hubble Space Telescope} (HST) image analysis.
In addition to this shape noise, there is a measurement error due to pixel noise. Assuming that the measurement error is independent of the shape noise, we produced the total ellipticity error as a sum of two Gaussian random numbers. The measurement error depends on galaxy properties (e.g., magnitude, size, profile shape, etc.) and signal-to-noise ratios (S/N).
We adopted the source magnitude distribution of the \citet{kim2019} study, which starts at $\mytilde21.5^{\rm th}$ mag, peaks at $\mytilde25.5^{\rm th}$ mag, and truncates at $\mytilde27.5^{\rm th}$ mag
in $V$-band.
The observed relation between magnitude and ellipticity measurement error in \citet{kim2019} is employed to 
generate the ellipticity measurement error for our sources.
We select 7,000 convergence fields and divide them into 5,000 training, 1,800 validation, and 200 test samples.

\subsubsection{Architecture of CNN} \label{sec:cnn_architecture}

\begin{figure*}[hbt]
\plotone{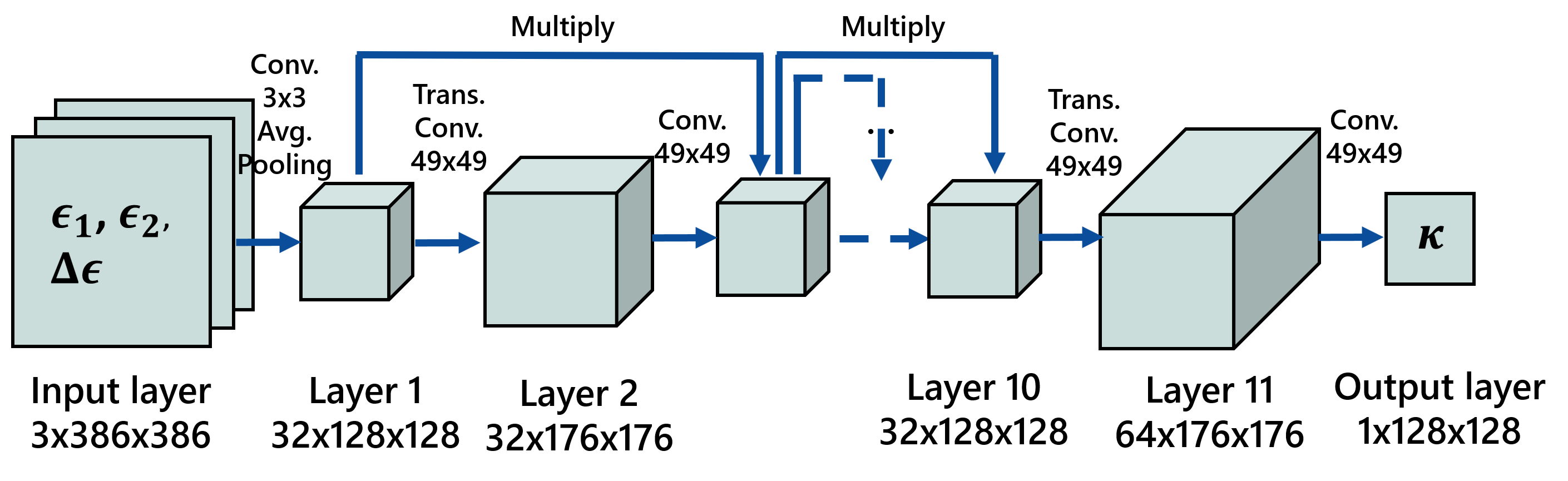}
\caption{Schematic diagram showing the architecture of our convolutional neural network.
The input channel consists of three ($\epsilon_1$, $\epsilon_2$, and $\Delta \epsilon$) layers of 2D ($386\times386$) arrays. The main part of the CNN architecture 
is the repeated combination of convolution and transposed-convolution with $49\times49$-size filters. We use skip-connections between the output of each convolution layer and that of the previous layer that has the matching size. See Table~\ref{tab:architecture} and text for details.}
\label{fig:architecture}
\end{figure*}

\begin{deluxetable}{cccccc}
\tablecaption{Outline of our convolutional neural network.
See text and Figure~\ref{fig:architecture} for description of each layer.\label{tab:architecture}
}
\tablehead {
\colhead{Layer} &
\colhead{Filter Size} &
\colhead{Multiplied to} &
\colhead{Output Size} 
}
\startdata
Input & - & - & (2, 386, 386) \\
Conv2D-1 & (3, 3) & - & (8, 384, 384) \\
AvgPool & (3, 3) & - & (8, 128, 128) \\
TransConv2D-1 & (49, 49) & - & (8, 176, 176) \\
Conv2D-2 & (49, 49) & - & (8, 128, 128) \\
Multiply-1 & - & AvgPool & (8, 128, 128) \\
TransConv2D-2 & (49, 49) & - & (8, 176, 176) \\
Conv2D-3 & (49, 49) & - & (8, 128, 128) \\
Multiply-2 & - & Multiply-1 & (8, 128, 128) \\
TransConv2D-3 & (49, 49) & - & (8, 176, 176) \\
Conv2D-4 & (49, 49) & - & (8, 128, 128) \\
Multiply-3 & - & Multiply-2 & (8, 128, 128) \\
TransConv2D-4 & (49, 49) & - & (8, 176, 176) \\
Conv2D-5 & (49, 49) & - & (8, 128, 128) \\
Multiply-4 & - & Multiply-1 & (8, 128, 128) \\
TransConv2D-5 & (49, 49) & - & (16, 176, 176) \\
Output & (49, 49) & - & (1, 128, 128) \\
\enddata
\end{deluxetable}

Figure~\ref{fig:architecture} and Table~\ref{tab:architecture} summarize the 
architecture of our CNN model that we use to predict the convergence map from the WL shear datasets.
Our CNN model takes two-dimensional (2D) arrays of $\epsilon_1 (\mathbf{x})$, $\epsilon_2 (\mathbf{x})$, and $\Delta \epsilon(\mathbf{x})$ as a three-channel input, where $\mathbf{x}$ denotes 2D pixel coordinates, $\epsilon_i (\mathbf{x}) (i=1,2)$ and $\Delta \epsilon(\mathbf{x})$ are the $i$-th average ellipticity (reduced shear) component and its error at the position $\mathbf{x}$. 
Because we randomly positioned source galaxies (\textsection\ref{sec:method_sim}), these (regularly spaced) input grids were constructed by weight-averaging the (irregularly spaced) source galaxy ellipticities with a FWHM$=7\arcsec$ Gaussian kernel; we used the distance between the center of each grid and the source position for the kernel evaluation.
For each cluster, the full area of the initial field is $32\arcmin\times32\arcmin$, which is represented by
2D arrays of $500\times500$. We performed data augmentation by subsampling $24\farcm7 \times 24\farcm7$ ($386\times386$) regions 1,444 times. In addition, we applied four rotations ($0^\circ$, $90^\circ$, $180^\circ$, and $270^\circ$) and two axis flips. The total number of the resulting subfields for each cluster is 1,444$\times(4+2)=$8,664. The subsampling scheme also prevents CNN from learning that the position of the cluster is always at the field center.

The main part of our CNN architecture includes the repeated combinations of 2D convolution (Conv2D-\# in Table~\ref{tab:architecture}) and transposed-convolution layers (TransConv2D-\#) with the identical filter size.
We tested various choices of filter sizes and found that the $49 \times 49$ filter gives the best overall performance.
Readers are referred to Appendix~\ref{sec:app_performance} for performance comparisons among different CNN architectures with various choices of filter sizes, input layers, and loss functions.
The Conv2D-\# and TransConv2D-\# operations are activated by the hyperbolic tangent function (tanh).
Inspired by the residual neural network \citep[ResNet;][]{he2015}, we use skip-connections between the output of each Conv2D-\# layer and that of the previous layer with the matching output size by applying a multiplication operation (Multiply-\#).
We apply batch normalization to the output of each Multiply-\# layer to avoid the so-called gradient vanishing problem \citep{ioffe2015}.
These repeated operations of Conv2D-\# and TransConv2D-\# are designed to extract features while preserving the size of the output layer (without introducing any padding).
Also, this architecture outperforms the other architectures that have only convolution layers when it comes to the prediction of the mass peak positions (Appendix~\ref{sec:app_performance}).
Although we did not use any arbitrary padding, the convergence estimates near the field boundary can easily be influenced by the nonvanishing filter size.
Therefore, the values within the 14 boundary pixels were not used during our training. The pixel scale of the final $\kappa$ map is $0\farcm192~\mbox{pixel}^{-1}$.

We performed our CNN training with Tensorflow \citep{tensorflow2015}.
During the training, we used the Adam optimizer \citep{adam2014} with a learning rate $10^{-5}$, $50$ mini-batches per each step, and $20$ steps per each epoch.
In this study, we introduce the following weighted mean square error inspired by the focal loss \citep{lin2017}:
\begin{equation}
\mathcal{L} = \sum_{\mathbf{x}} \omega_{\rm f}(\mathbf{x}) \left[ \kappa_{\rm pred}(\mathbf{x}) - \kappa_{\rm truth}(\mathbf{x}) \right]^2 \, ,
\label{eq:focal_loss}
\end{equation}
where the weight $\omega_{\rm f}(\mathbf{x})$ at each pixel is determined by the value of truth convergence:
\begin{equation}
\omega_{\rm f}(\mathbf{x}) = 1 + \frac{\left| \kappa_{\rm truth}(\mathbf{x}) \right|}{\max (\kappa_{\rm truth})} \, .
\end{equation}
This loss function is chosen so that our CNN model is mostly constrained by the high-density regions of clusters, where our scientific interests lie (see Appendix~\ref{sec:app_performance} for performance comparisons).
We ran $200$ epochs with the NVIDIA V100 GPU, which take about one hour per each training.
For the convergence test of our CNN training, we executed 10 independent runs with the same CNN architecture.
For each run, we 
adopted the model that minimized the validation loss.
In our presentation of the results (\textsection\ref{sec:results}), the standard deviations from the 10 runs are used as error estimates.


\section{Results}\label{sec:results}

In this section, we compare our CNN 
results
with those of KS93 for the test sample comprised of 200 cluster fields.
Because of the data augmentation procedure (\textsection\ref{sec:method_cnn}), multiple (subsampled) mass maps are produced for each cluster. Thus, we created one mosaic convergence image for each cluster by taking average of the multiple mass maps. The cluster is approximately located at the center in this mosaic and we use the mosaic for comparison with the truth and KS93 results.
Readers are reminded that since these multiple mass maps are generated from the same source catalog, this mosaicking procedure does not benefit by reducing the statistical noise\footnote{That is, one can apply the procedure to real observations.}.
In \textsection\ref{sec:results_visual}, we use visual inspection to qualitatively compare the  
reconstruction results.
In \textsection\ref{sec:results_prob}, we contrast the values of reconstructed convergence with those of
the truth 
 by pixel-by-pixel comparison and by evaluating their probability distributions.
In \textsection\ref{sec:results_mass} and \textsection\ref{sec:results_peak}, we 
investigate the reconstructed cluster masses and the positional accuracy of their density peaks, respectively.
Finally, we examine performances of our CNN method in the presence of bright  stars in \textsection\ref{sec:results_star}.
Table~\ref{tab:performance} summarizes our comparison between the KS93 and CNN results.

\begin{deluxetable}{cccc}[hbt]
\tablecaption{Summary of the 
performances of our CNN and the KS93 mass reconstructions for the 200 test datasets.
See text for the definition of the $\mathcal {D}$ metric.}\label{tab:performance}
\tablehead {
\colhead{Method} &
\colhead{$\mathcal{D}(\widetilde{\kappa}_{\rm pred},\widetilde{\kappa}_{\rm truth})$} &
\colhead{$M_{\rm pred}^{\rm cl}/M_{\rm truth}^{\rm cl}$} &
\colhead{$\Delta_{\rm peak}$}
}
\startdata
KS93 & $6.26 \pm 4.57$ & $0.484^{+0.218}_{-0.149}$ & 
$4\farcm27^{+8\farcm63}_{-3\farcm94}$ \\
CNN & $4.36 \pm 3.77$ & ${0.867^{+0.327}_{-0.296}}$ & 
$0\farcm60^{+4\farcm92}_{-0\farcm38}$ \\
CNN-BS & ${3.66 \pm 3.22}$ & ${0.554^{+0.257}_{-0.211}}$ &
$1\farcm53^{+4\farcm21}_{-1\farcm14}$ \\
\enddata
\end{deluxetable}


\subsection{Qualitative Comparison Based on Visual Inspection}\label{sec:results_visual}

\begin{figure*}[hbt]
\plotone{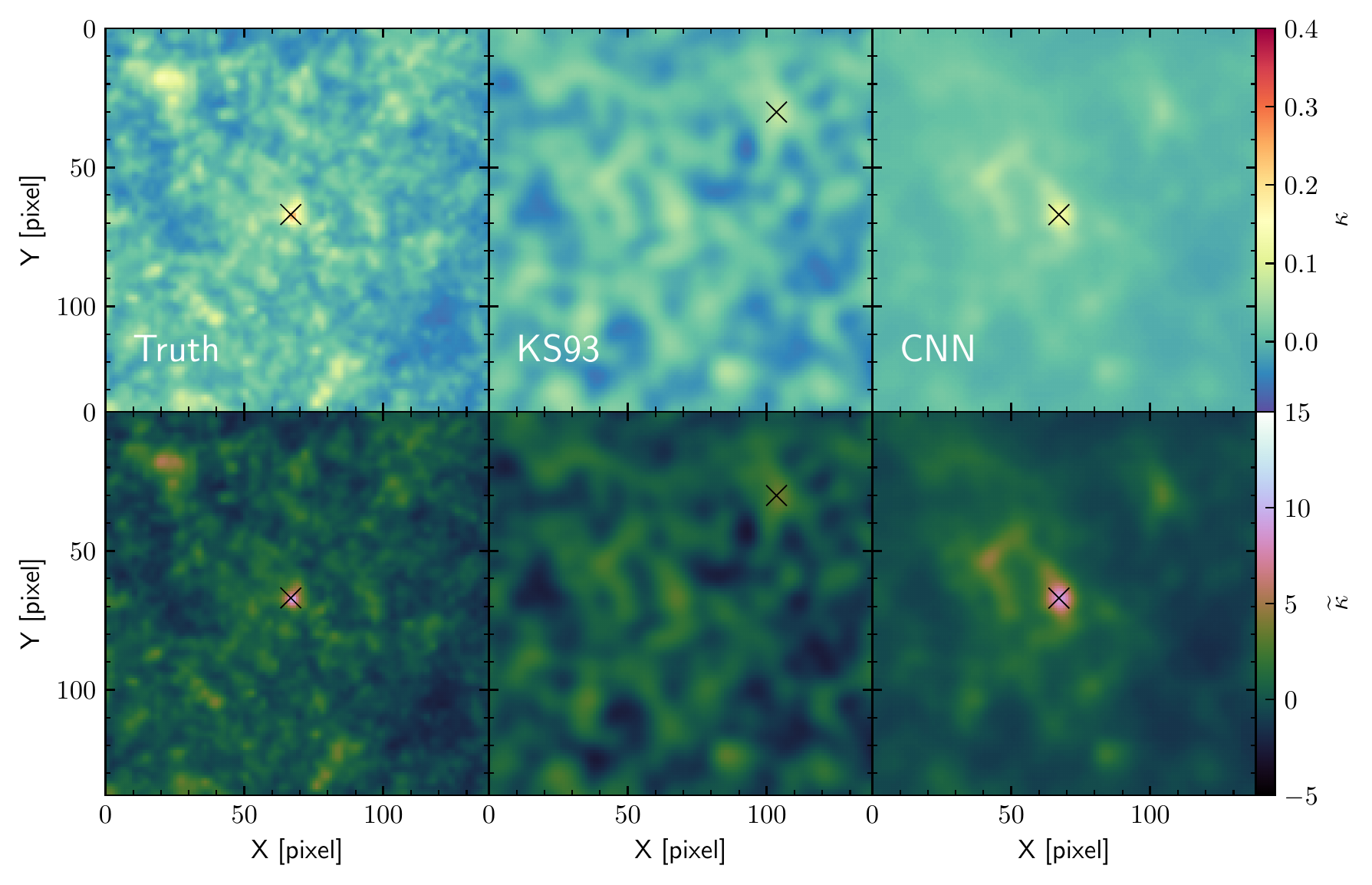}
\caption{Example of our CNN mass reconstruction. We show the truth convergence map (left) and the reconstructions with the KS93 (middle) and our CNN methods (right).
The top panel displays the convergence values $\kappa(\mathbf{x})$ as are whereas the bottom panel shows the rescaled versions using the transformation
$\widetilde{\kappa}(\mathbf{x}) \equiv (\kappa(\mathbf{x}) - \langle \kappa \rangle)  / \Delta \kappa$, where $\langle \kappa \rangle$ and $\Delta \kappa$ are the average and standard deviation, respectively, evaluated within the field.
The ``X" symbol denotes the location of the highest value within each convergence field. 
Here we only display the central $26\farcm6 \times 26\farcm6$ region.
Visual inspection shows that our CNN reconstruction significantly outperforms the KS93 method in terms of the dynamical range restoration, noise suppression, and large-scale structure representation.}
\label{fig:visual_sim}
\end{figure*}

Figure~\ref{fig:visual_sim} displays an example of our CNN mass reconstruction. The comparison with the truth and KS93 results illustrates that our CNN mass reconstruction is superior to KS93 in terms of 1) the recovery of the true $\kappa$ range, 2) the representation of the large-scale structure around the cluster, and 3) the suppression of the noise in the cluster outskirts.
Although here only one case is illustrated, these advantages are present for the rest of the test sample.

\begin{description}
\item[1) Recovery of the $\kappa$ range] The truth map shows that $\kappa$ ranges from $\mytilde0.05$ to $\mytilde0.25$, where the maximum value is found at the cluster center. The KS93 reconstruction fails to recover this convergence range in the high end. The convergence value at the cluster center is only $\mytilde0.05$ while the global maximum is found at a different location (see the location of the ``X" symbol). On the other hand, the CNN mass reconstruction gives a much higher value $\kappa\sim0.15$ at the cluster center. Given the inevitable smoothing effect arising from the sparse sampling (25 sources per sq. arcmin), we believe that the improvement over the KS93 is remarkable. We discuss this issue more quantitatively in \textsection\ref{sec:results_prob}.
\item [2) Reconstruction of the large scale structure] Inspection of the truth map (Figure~\ref{fig:visual_sim}) indicates that the cluster is not isolated, but located in the high density environment. While it is difficult to trace this large-scale structure surrounding the cluster in the KS93 result, the feature, albeit somewhat smoothed, clearly stands out as an overdense region in our CNN mass reconstruction. 
\item [3) Suppression of noise] Since the S/N value depends on the local strength of WL signal given the same number density of sources, an ideal mass reconstruction method should employ a  smoothing scheme where the kernel size matches the local S/N value. However, in general, it is nontrivial to implement such an ``adaptive smoothing" scheme in practice because the S/N information is only obtained after a high-quality convergence field is reconstructed. Therefore, a common practice in the WL community is to perform mass reconstruction with a fixed-size smoothing kernel often optimized for the central region of the cluster \citep[e.g.,][]{vanwaerberke2000}. The 
inevitable artifact is the production of many spurious mass peaks in the cluster outskirt where the S/N value is low. The comparison between the KS93 and our CNN results shows that our CNN result nicely suppresses the noise fluctuation in the outskirt region while still detecting  substructures if they are significant (see the bottom panel of Figure~\ref{fig:visual_sim}).
\end{description}


\subsection{Convergence Distribution}\label{sec:results_prob}

\begin{figure*}[hbt]
\plottwo{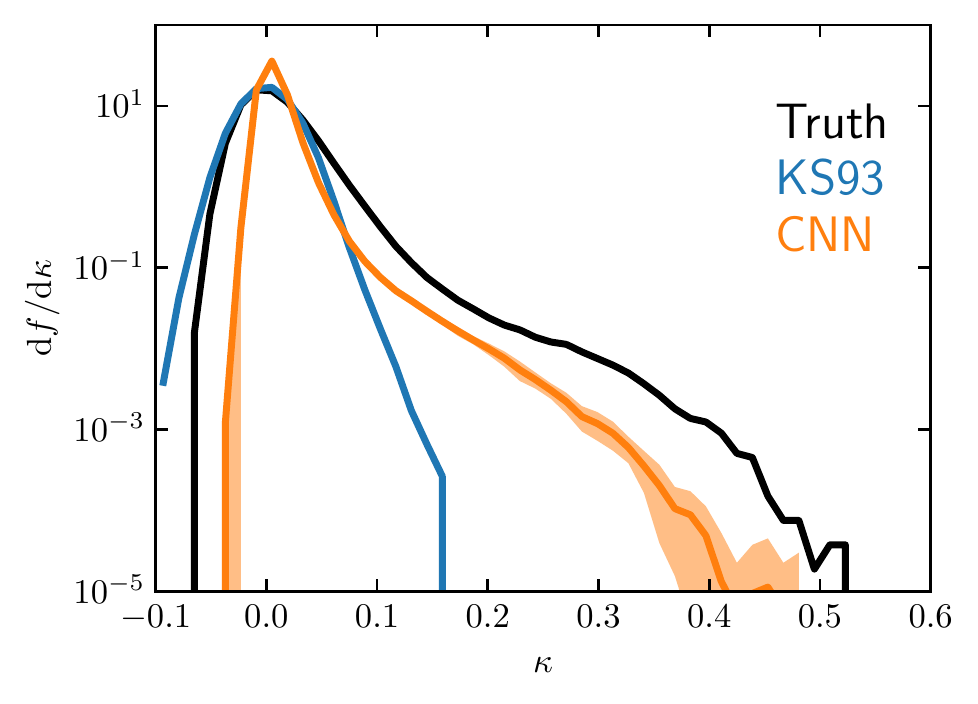}{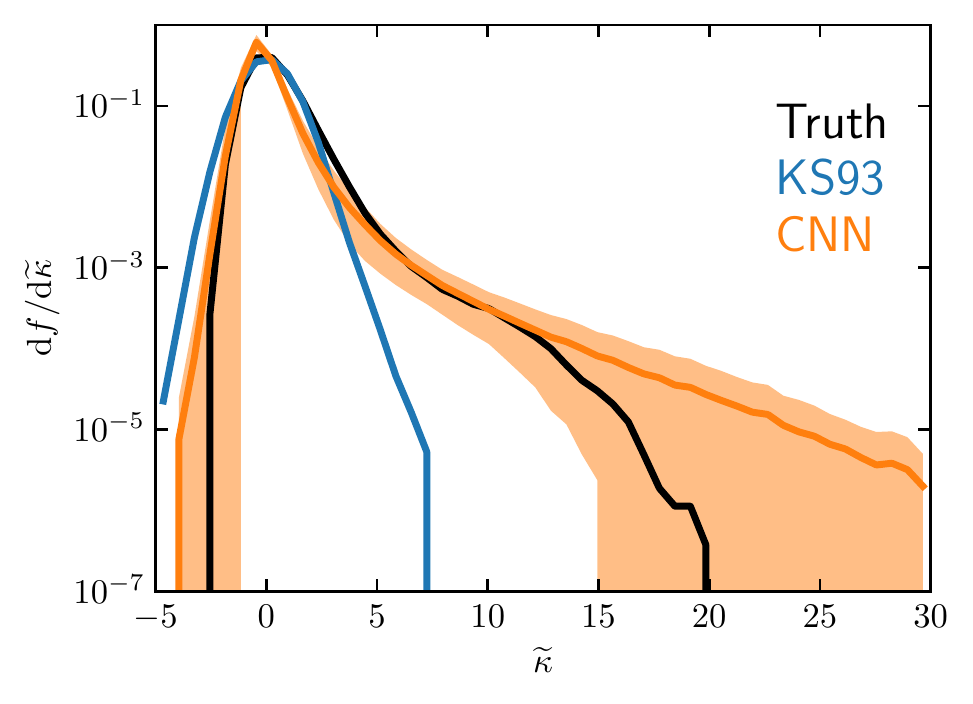}
\caption{Comparison of convergence ($\kappa$) distributions. 
We measure the $\kappa$ distribution from the entire test sample (left). The right panel is the same except that the distribution is obtained for $\widetilde{\kappa}$.
The orange shade represents the standard deviation measured from our 10 independent runs.
The CNN reconstruction provides an extended tail at the high end, mimicking the feature in the truth whereas the KS93 distribution is nearly symmetric around zero. When the convergence is rescaled with its standard deviation, the agreement improves as can be seen in the right panel.}
\label{fig:kappaHist}
\end{figure*}

\begin{figure*}[hbt]
\centering
\plottwo{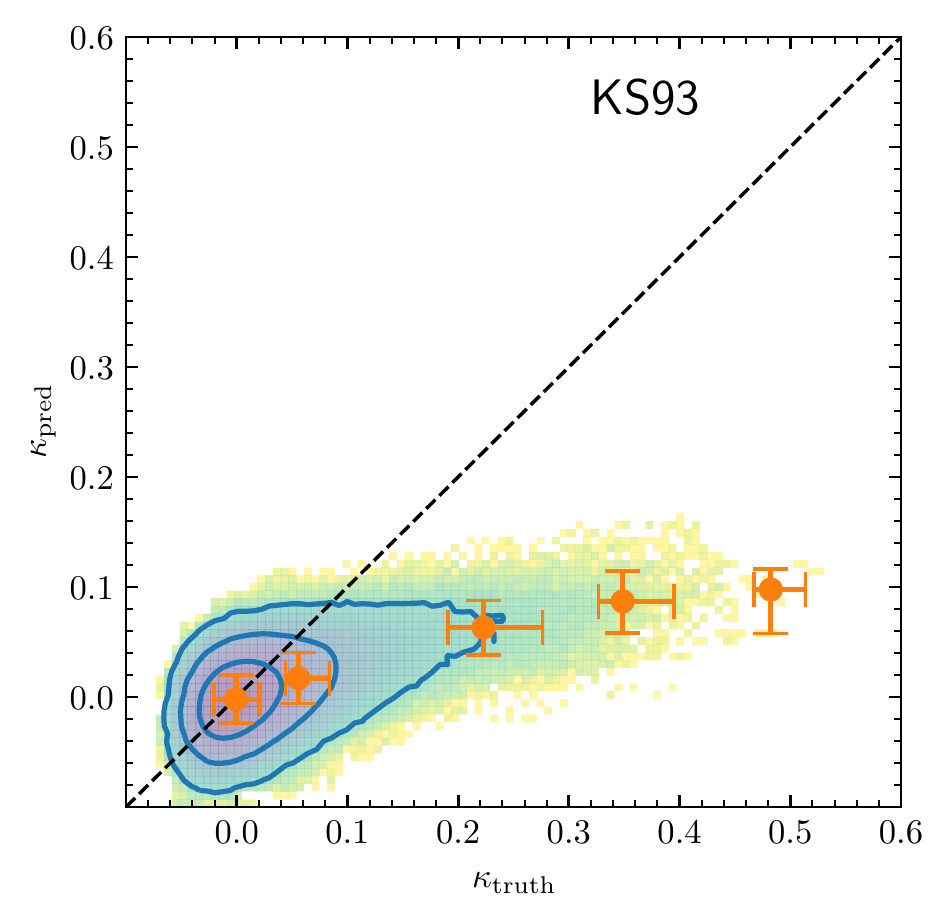}{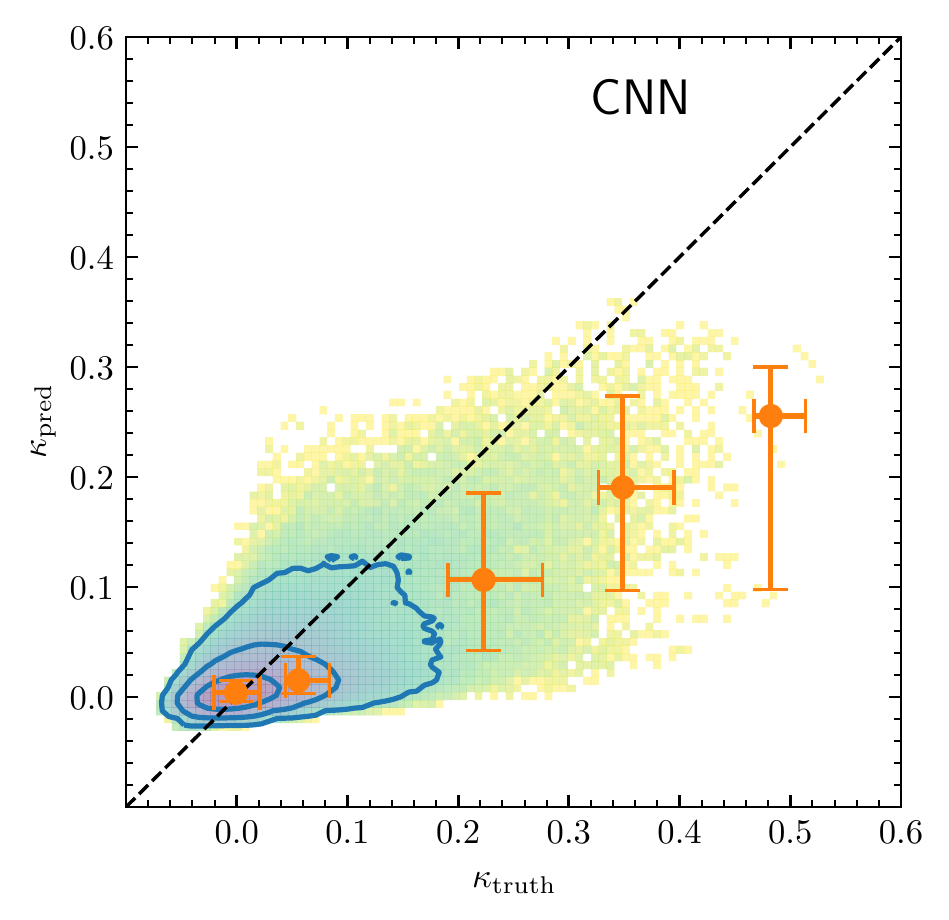}
\caption{Joint distribution of $\kappa$ between reconstructed and truth convergence fields.
Contours show the $68\%$, $95\%$, and $99.7\%$
confidence levels.
Orange filled circles are the medians of $\kappa_{\rm truth}$ in the equal-width
bins and their error bars represent the $68\%$ 
certainties in $\kappa_{\rm truth}$ and $\kappa_{\rm pred}$.
The CNN reconstruction shows an improved correlation with the truth, although it is also clear that the $\kappa$ values are underestimated.}
\label{fig:pixel}
\end{figure*}

\begin{figure*}[hbt]
\plottwo{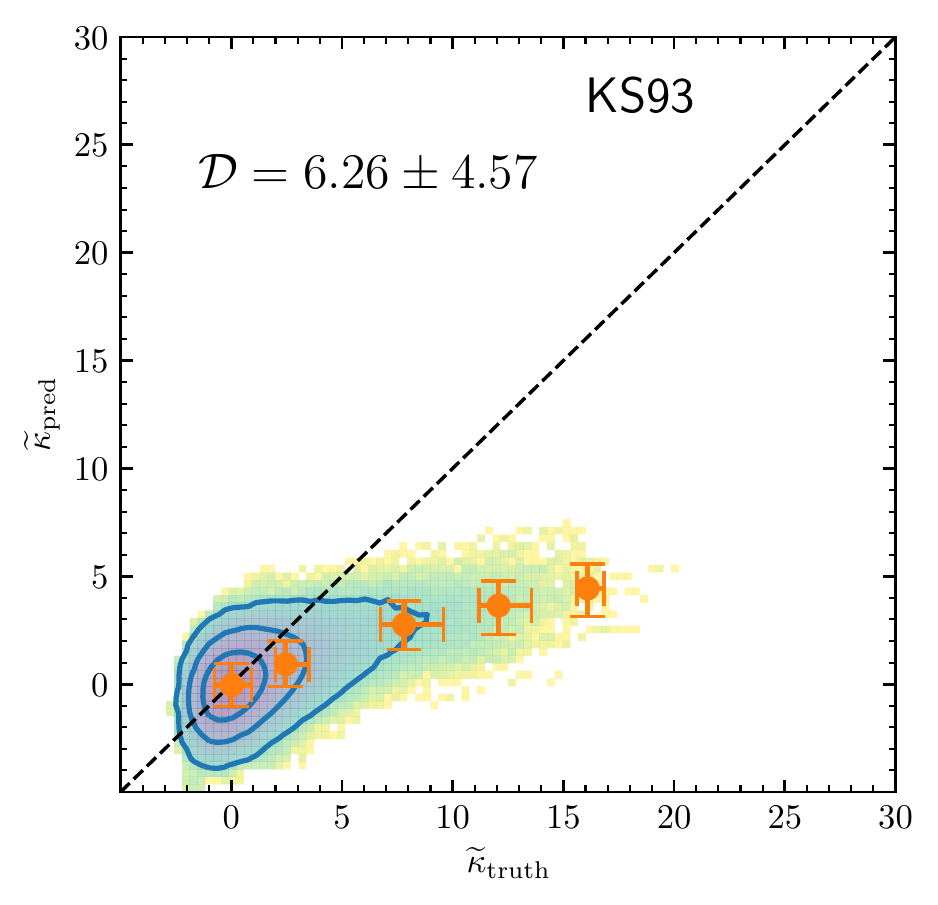}{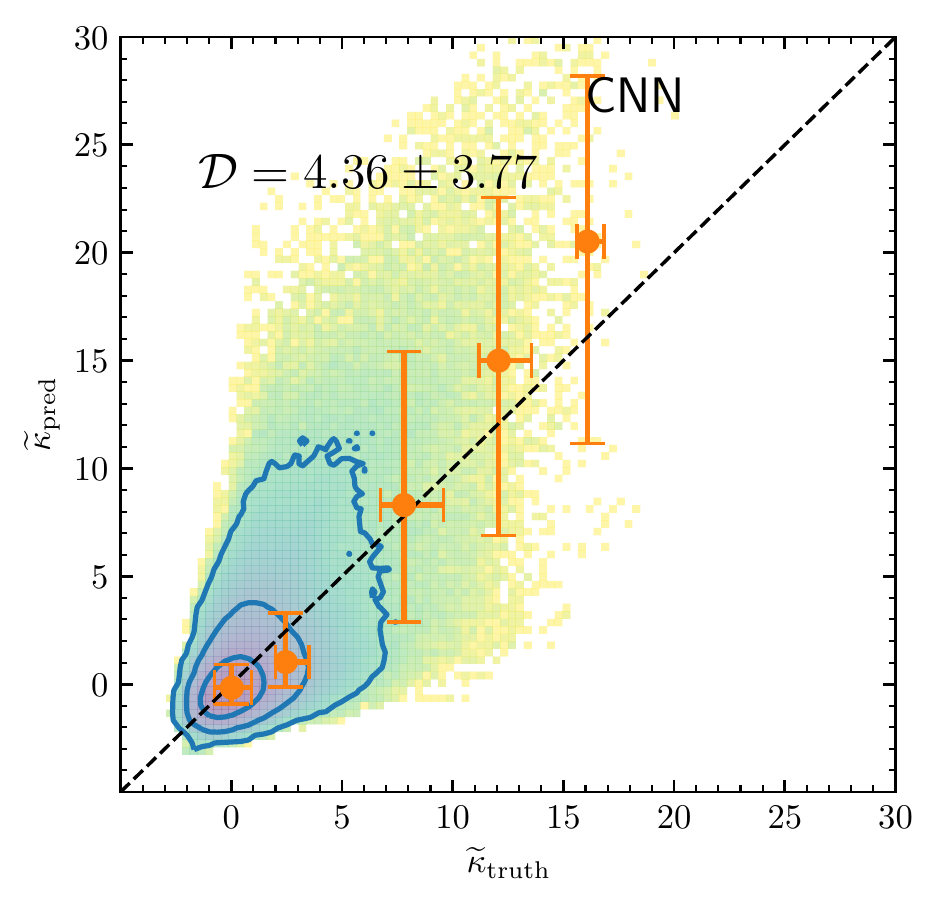}
\caption{Similar to 
Figure~\ref{fig:pixel} except that the plots are drawn with the normalized convergence ($\widetilde{\kappa}$).}
\label{fig:pixelNorm}
\end{figure*}

The literature has shown that the distribution of the convergence field can be well-approximated by a log-normal distribution characterized by an extended high-end tail \citep[e.g.,][]{jain2000,hilbert2011,clerkin2017}.
Figure~\ref{fig:kappaHist} compares the convergence distributions between the KS93 and our CNN reconstructions for the entire test sample and shows
that the CNN distribution follows the log-normal trend of the truth (see the left panel). On the other hand, the convergence distribution in the KS93 result is symmetric around zero without any sign of an extended tail at the high end.

The comparison between the CNN result and the truth shows that the CNN distribution is somewhat narrower. This happens because the CNN mass map based on a finite number of source galaxies (25 galaxies per sq. arcmin) is inevitably smoother than the truth. In order to compensate for this smoothing effect, we propose the following normalization:
\begin{equation}
\widetilde{\kappa}(\mathbf{x}) \equiv (\kappa(\mathbf{x}) - \langle \kappa \rangle)  / \Delta \kappa, \label{eqn:normalization}
\end{equation}
where $\langle \kappa \rangle$ and $\Delta \kappa$ are the average and standard deviation, respectively.
The rescaling of the convergence through this normalization takes into account the reduction of $\Delta \kappa$ in mass reconstruction. However, as mentioned in \textsection\ref{sec:results_visual}, the smoothing kernel is not uniform in the CNN mass reconstruction; effectively, the cluster outskirts smoothed with larger kernels than the cores. Therefore, the proposed normalization (equation~\ref{eqn:normalization}) does not completely resolve the issue.
Nevertheless, the right panel of Figure~\ref{fig:kappaHist} shows that the agreement between the CNN and truth distributions improves dramatically after the normalization. 

The joint distribution shown in Figure~\ref{fig:pixel} also confirms that the CNN mass reconstruction provides significantly better pixel-to-pixel correlations with the truth. Also, similarly to the previous case, the normalization significantly strengthens the correlation with the truth (Figure~\ref{fig:pixelNorm}).

To quantify the similarity of the reconstruction to the truth, one can suggest the absolute deviation
$|\widetilde{\kappa}_{\rm pred}(\mathbf{x}) - \widetilde{\kappa}_{\rm truth}(\mathbf{x})|$ as a potential metric. However, this metric, if used as it is, would be dominated by the statistics of the convergence pixels near zero. Therefore, we introduce the weighted version as follows:
\begin{equation}
\mathcal{D} = \frac{\sum_{\mathbf{x}} \widetilde{\omega}_{\rm p}(\mathbf{x}) \left| \widetilde{\kappa}_{\rm pred}(\mathbf{x}) - \widetilde{\kappa}_{\rm truth}(\mathbf{x}) \right|}{\sum_{\mathbf{x}} \widetilde{\omega}_{\rm p}(\mathbf{x})}\label{eqn:dev},
\end{equation}
where the weight $\widetilde{\omega}_{\rm p}(\mathbf{x})$ is inversely proportional to the probability distribution:
\begin{equation}
\frac{1}{\widetilde{\omega}_{\rm p}(\mathbf{x})} = \left. \frac{{\rm d}f}{{\rm d}\widetilde{\kappa}_{\rm truth}} \right|_{\widetilde{\kappa}_{\rm truth}(\mathbf{x})} .
\end{equation}
In practice, the weight can diverge when noise makes the derivatives  close to zero. To prevent this, we use a discrete histogram of the truth with 50 bins for the estimation of $\widetilde{\omega}_{\rm p}$.
The $\mathcal{D}$ (better if closer to zero)
metric from the CNN mass reconstruction is
$4.36 \pm 3.77$. On the other hand,
the KS93 result gives $\mathcal{D}=6.26 \pm 4.57$.
This metric indicates that the $\kappa$ statistics from the CNN reconstruction better match those from the truth.


\subsection{Projected Cluster Mass}\label{sec:results_mass}

\begin{figure*}[hbt]
\plottwo{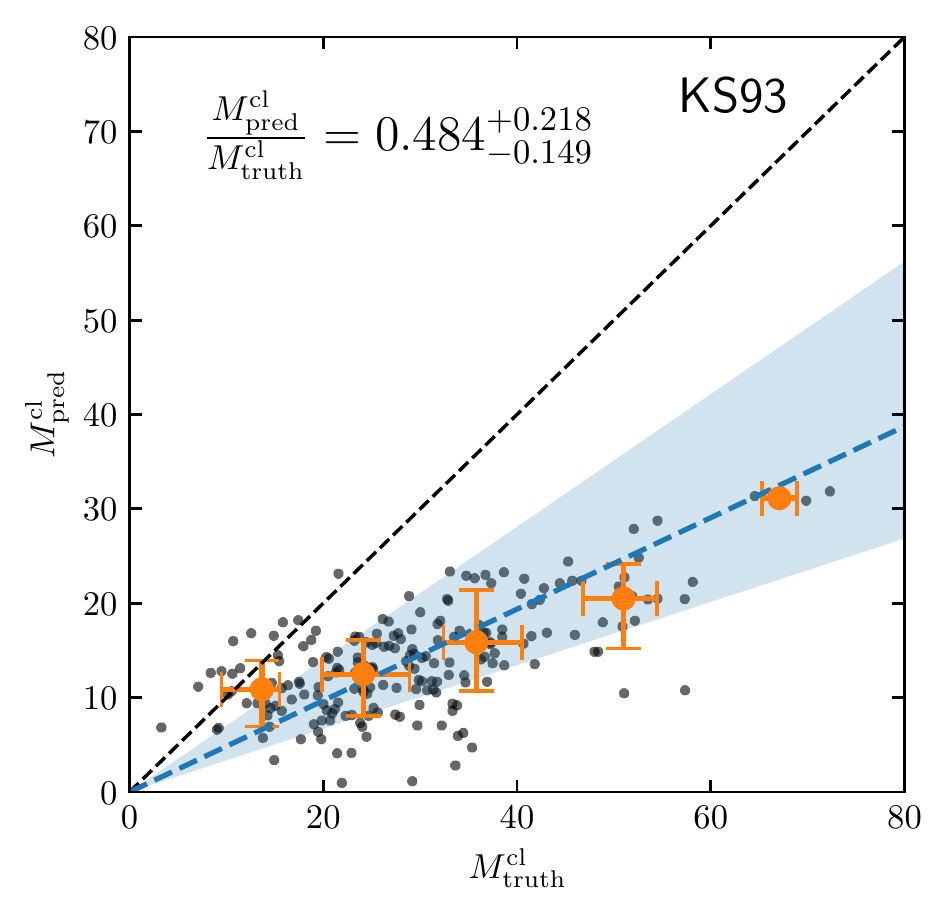}{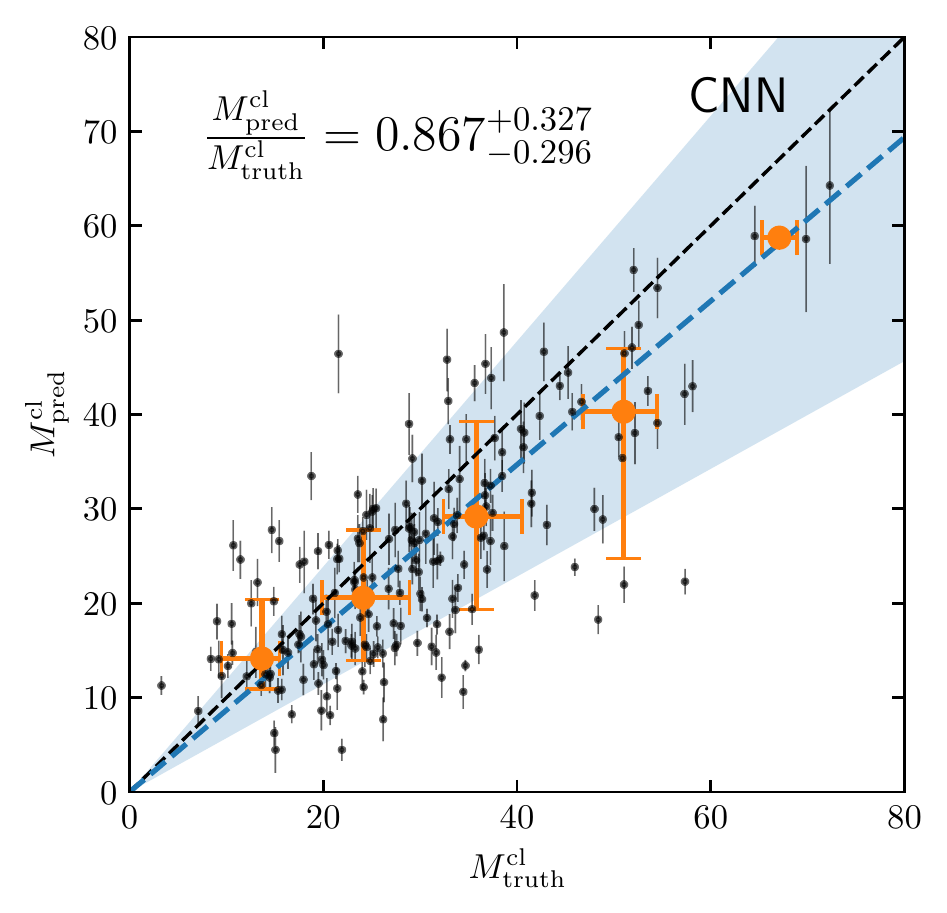}
\caption{Comparison of projected cluster masses between prediction and truth. The projected cluster mass $M^{\rm cl}$ is defined to be the sum of convergence values
within a $r=1\farcm92$ ($\mytilde0.72$ Mpc) radius circular aperture centered on the truth mass peak.
Filled orange circles are the median values of $M_{\rm truth}^{\rm cl}$ in the equal-width
bins, and their error bars represent the
$68\%$
certainties of $M_{\rm truth}^{\rm cl}$ ad $M_{\rm pred}^{\rm cl}$ within the bins.
The errors on individual data points (filled black circles) in the right panel are the standard deviations of CNN results from 10 independent runs. 
Blue dashed lines and filled area are the median and $68\%$
confidence levels of the ratio $M_{\rm pred}^{\rm cl} / M_{\rm truth}^{\rm cl}$.
The projected mass of $M^{\rm cl}=10$ approximately corresponds to $M_{200}\sim7\times10^{13} M_{\sun}$.
The ratio of the CNN masses to the truth is consistent with unity ($0.867_{-0.296}^{+0.327}$) while the ratio is significantly lower 
($0.484_{-0.149}^{+0.218}$) when the KS93 masses are used.}
\label{fig:mass}
\end{figure*}

The pixel-to-pixel comparison in \textsection\ref{sec:results_prob} shows that although our CNN mass reconstruction better recovers the convergence statistics of the truth than the KS93 method, the distribution is somewhat narrower because of the smoothing implicitly applied to the reconstructed convergence field via CNN. It is our premise that this smoothing artifact is of a less concern when one's interest is to estimate the integrated convergence within a reasonably large aperture. We define
the projected cluster mass $M_{\rm truth}^{\rm cl}$ to be the sum
of the convergence values within the $r=1\farcm92$ (10 convergence pixel) radius aperture. 
At the cluster redshift of 0.5, the radius corresponds to 0.72~Mpc with the adopted cosmology. A projected cluster mass of 10 ($\sum \kappa=10$) corresponds to $\mytilde7\times10^{13}M_{\sun}$\footnote{The exact value depends on the halo profile shape. Here we assume an NFW profile with a concentration of $c=3.5$ and a scale radius of $r_s=200$~kpc at $z=0.5$.}.

Figure~\ref{fig:mass} shows the comparison of $M_{\rm truth}^{\rm cl}$  between the reconstructed and the truth values.
As seen in the pixel-to-pixel comparison, the CNN mass reconstruction also outperforms the KS93 result in the cluster mass estimation. In addition, it is remarkable that the agreement with the truth is significantly better than the one in the convergence pixel-to-pixel comparison. 
The slope $M_{\rm CNN}^{\rm cl} / M_{\rm truth}^{\rm cl} = 0.867^{+0.327}_{-0.296}$ is consistent with unity.
On the other hand, we obtain $M_{\rm KS}^{\rm cl} / M_{\rm truth}^{\rm cl} = 0.484^{+0.218}_{-0.149}$ for the KS93 reconstruction, which is a $\gtrsim2\sigma$ departure from unity.
For the case of CNN, the data points at $M_{\rm truth}^{\rm cl}\gtrsim40$ hint at the possibility that the estimated masses may be systematically lower. Although the sample size is small in this regime, we speculate that this may happen because the employed aperture radius ($r=1\farcm92$) is not sufficiently large for these very massive clusters.


\subsection{Cluster Centroid}\label{sec:results_peak} 

\begin{figure*}[hbt]
\plottwo{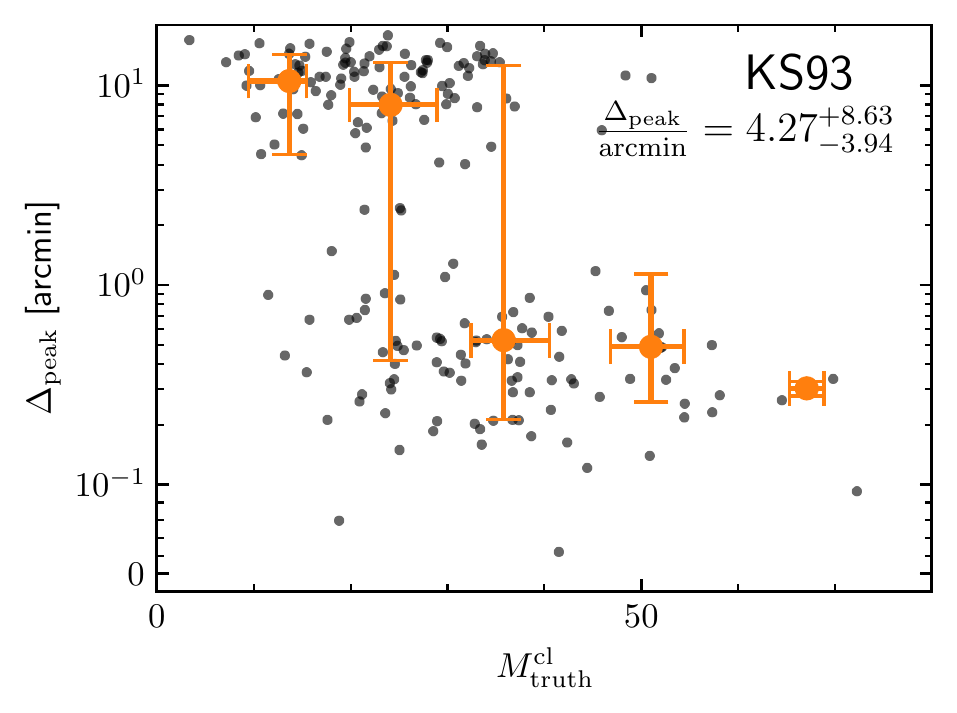}{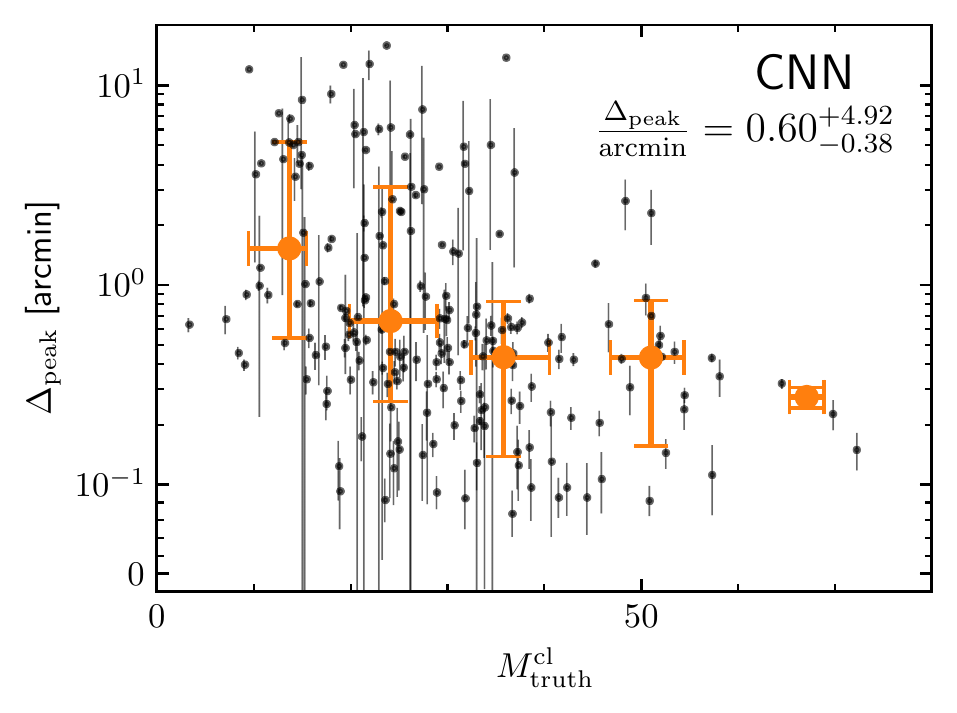}
\caption{Cluster centroid deviation
($\Delta_{\rm peak}$) as a function of the truth cluster mass ($M_{\rm truth}^{\rm cl}$).
Both CNN and KS93 perform well for massive clusters ($M_{\rm truth}^{\rm cl}\gtrsim35$). However, the KS93 method produces many catastrophic errors for $M_{\rm truth}^{\rm cl}\lesssim35$.
}
\label{fig:peakDistMass}
\end{figure*}

\begin{figure}[hbt]
\plotone{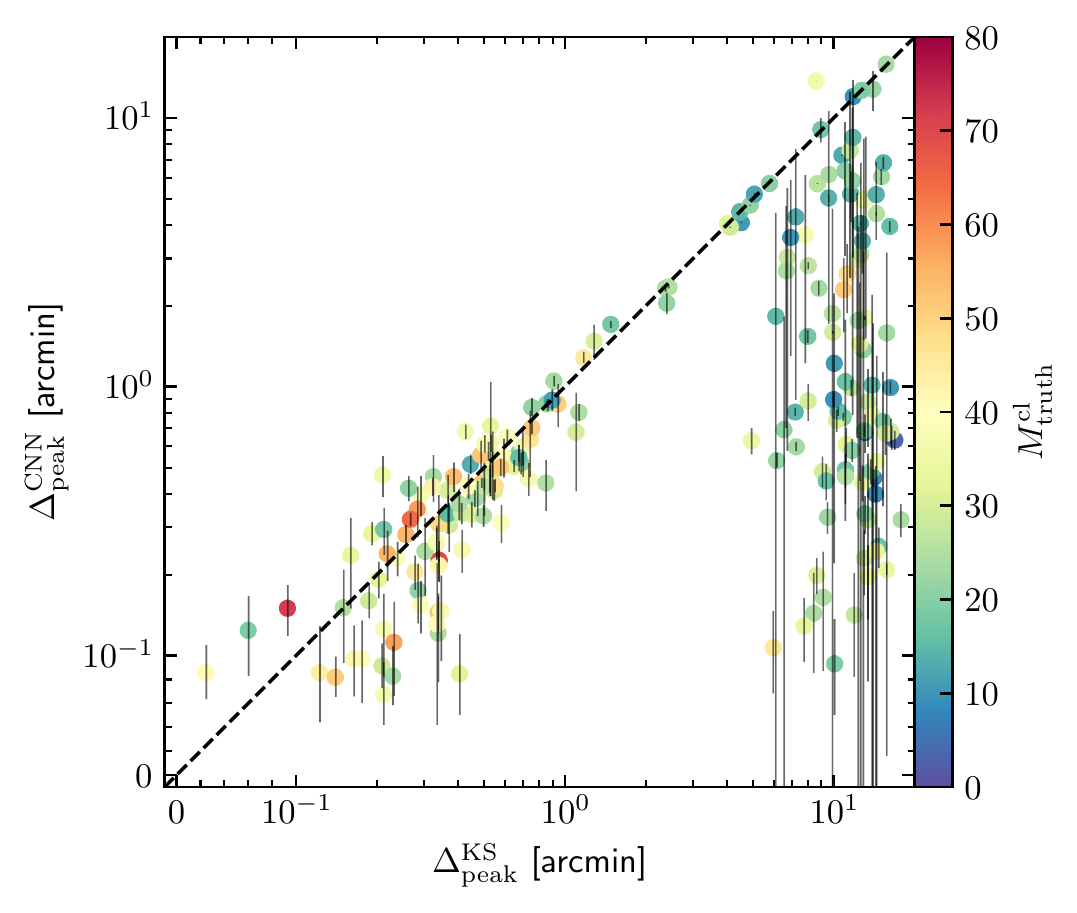}
\caption{Comparison of the centroid errors
between the KS93 and CNN results.
The data points are color-coded with the truth mass.
There present many catastrophic errors in the KS93 results.
}
\label{fig:peakDist}
\end{figure}

Robust estimation of centroids is an important issue in cluster WL studies \citep[e.g.,][]{linden2014,randall2008}. The centroid serves as a reference to characterize the properties of the cluster. Also, in merging galaxy clusters, the position of the mass clump with respect to other cluster components is critical in our reconstruction of their merging scenarios. Here we compare the performance in centroid recovery between the KS93 and our CNN methods.

We measured the centroid in two steps. First, we located the pixel that has the
largest convergence value. Then, we applied a 21~pixel$\times$21~pixel ($4\farcm03\times4\farcm03$)
square top-hat window and evaluated the first moments. 
Occasionally, negative convergence values are present within the window in the KS93 mass reconstruction. To prevent the centroid from leaving the window in this case, we rescaled the mass map in such a way that the minimum value within the window becomes zero.
The application of the top-hat window is to 
include the contribution from the large-scale structure around the peak in our estimation of the centroid. 

Figures~\ref{fig:peakDistMass} displays the deviations of the reconstructed mass centroids with respect to the truth.
The CNN and KS93 results give similarly small ($1 \sim 3$ pixels) centroid deviations for massive clusters ($M_{\rm truth}^{\rm cl}\gtrsim35$). Remarkably, we find striking differences in the low-mass ($M_{\rm truth}^{\rm cl}\lesssim35$)
regime. The CNN centroid deviations gradually increase for decreasing masses, reaching $\mytilde10$~pixels at $M_{\rm truth}^{\rm cl}\sim10$. On the other hand, the KS93 result shows many catastrophic errors ($\gtrsim50$ pixels) in this regime.
This contrast is seen more clearly in Figure~\ref{fig:peakDist}, where we directly compare the deviations for the same clusters.

We attribute the large difference in the centroid deviations for low mass clusters to the uncontrolled noise fluctuation in the KS93 mass reconstruction discussed in \textsection\ref{sec:results_visual}. As shown by the example in Figure~\ref{fig:visual_sim}, sometimes the highest convergence values are found not within the cluster region. Also, even in the case where the highest convergence value is not catastrophically far from the truth, the lack of the contrast against the neighboring background substructures makes the centroid measurement highly uncertain.


\subsection{Influence of Masking}\label{sec:results_star}

Up to now, we have tested our CNN method 
while assuming that no masked regions are present within the reconstruction field.
In real observations, however, we need to mask out the regions affected by bright stars.
Several methods have been suggested to minimize some artifacts due to the missing information
\citep[e.g.,][]{starck2003,pires2009}. In this paper, without taking any explicit measure to minimize the influence
(i.e., we did not perform a separate training with masked galaxy catalogs), we simply investigated the impact of large maskings on our CNN mass reconstruction performance with the same model.

The expected number density of bright stars depends on the galactic latitude. And the exact size of the masking for a given magnitude star varies according to specific reduction/analysis methods. Reviewing our previous WL studies with Subaru/Suprime-Cam imaging data, we find that 
within the typical $20\arcmin \times 20\arcmin$ WL analysis area, $1\sim2$ bright-star maskings were needed with the masking radius ranging from $\mytilde0.5\arcmin$ to $\mytilde2\arcmin$ \citep[e.g.,][]{finner2017, kim2019, yoon2020}.
To mimic such conditions, we applied bright-star masking to our source catalogs with these number density and size distributions. We ensured that every cluster has at least one masking near the mass peak because we are interested in investigating the effect at its maximum.

\begin{figure*}
\plotone{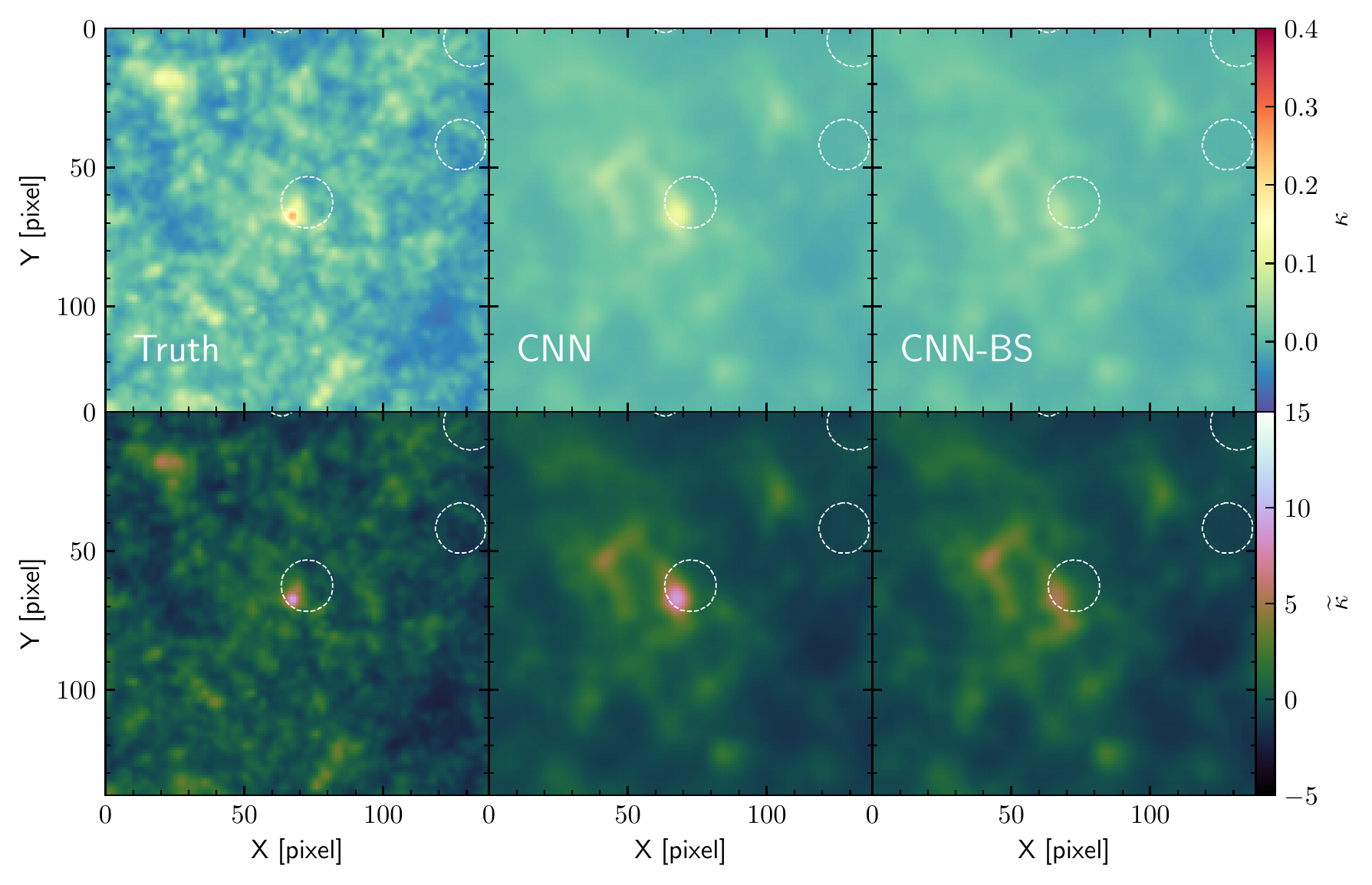}
\caption{Impact of the bright-star masking on the CNN mass reconstruction. We use the same cluster presented in Figure~\ref{fig:visual_sim} as an example.
The middle (right) panel shows the reconstruction without (with) bright-star masking. 
The color scale of the bottom panel is based on the normalized convergence ($\widetilde{\kappa}$) as in Figure~\ref{fig:visual_sim}.
White dashed circles mark the locations of the bright-star masks.
Although the central mask crops out the most significant region of the cluster, the result (CNN-BS) shows that the cluster is still clearly detected near the truth position. We note that the missing information leads to slight underestimation of the peak convergence values.
}

\label{fig:visual_BS}
\end{figure*}

Our visual inspection of the result shows that in most cases the CNN method can still detect the cluster mass clumps in the presence of masks. In Figure~\ref{fig:visual_BS} we display one such example. Although the central mask completely removes source galaxies 
within the $r=2\arcmin$ ($\mytilde0.73$ Mpc) circular mask placed near the mass peak, the reconstruction can still reveal the cluster nearly at the truth position. However, we find that because of the missing data the convergence values are slightly underestimated.

\begin{figure*}[hbt]
\centering
\includegraphics[width=0.3\textwidth]{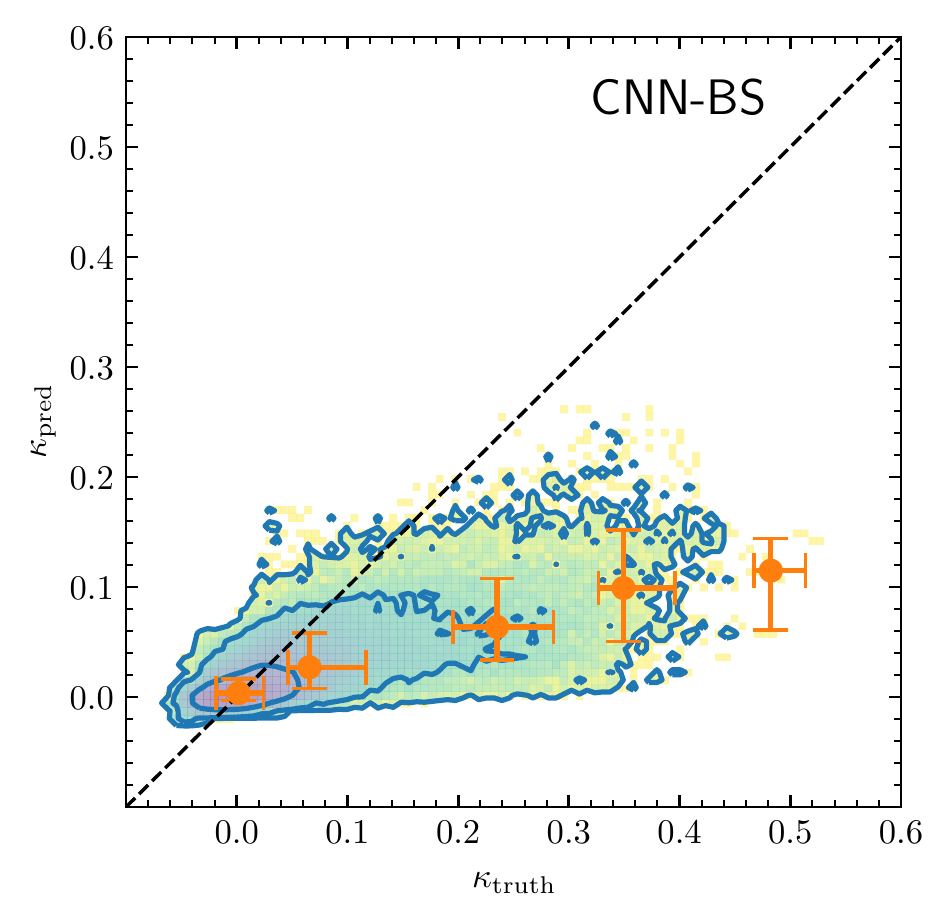}
\includegraphics[width=0.3\textwidth]{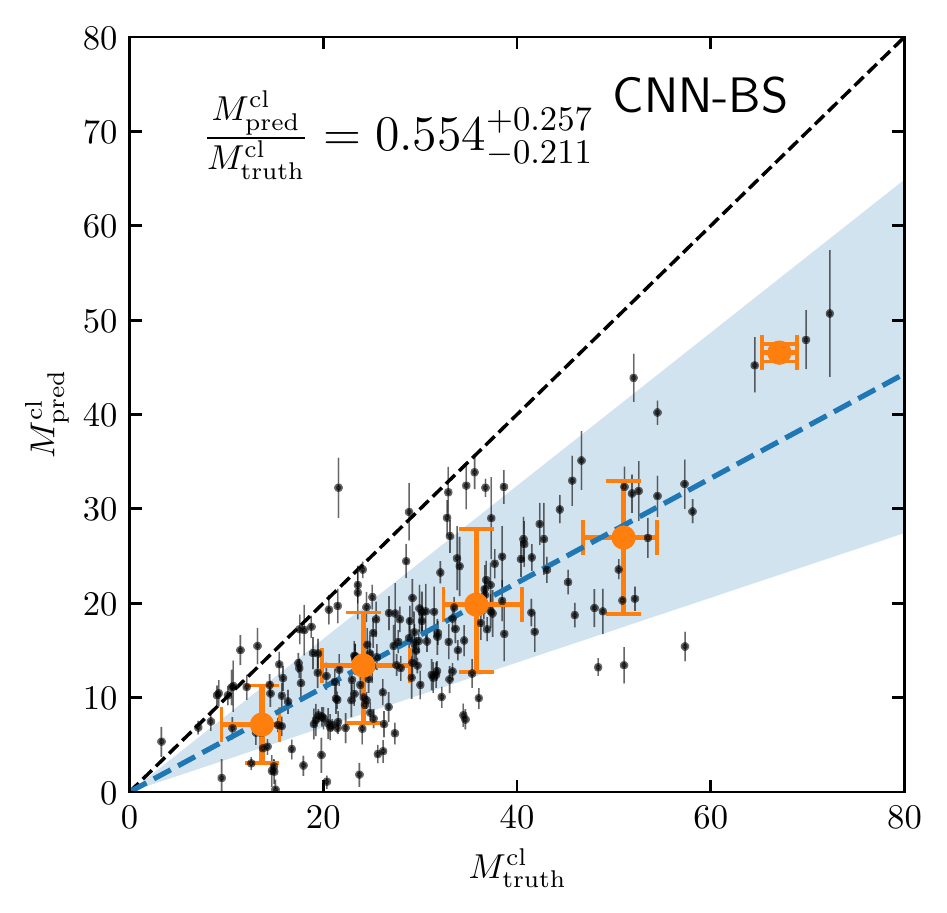}
\includegraphics[width=0.33\textwidth]{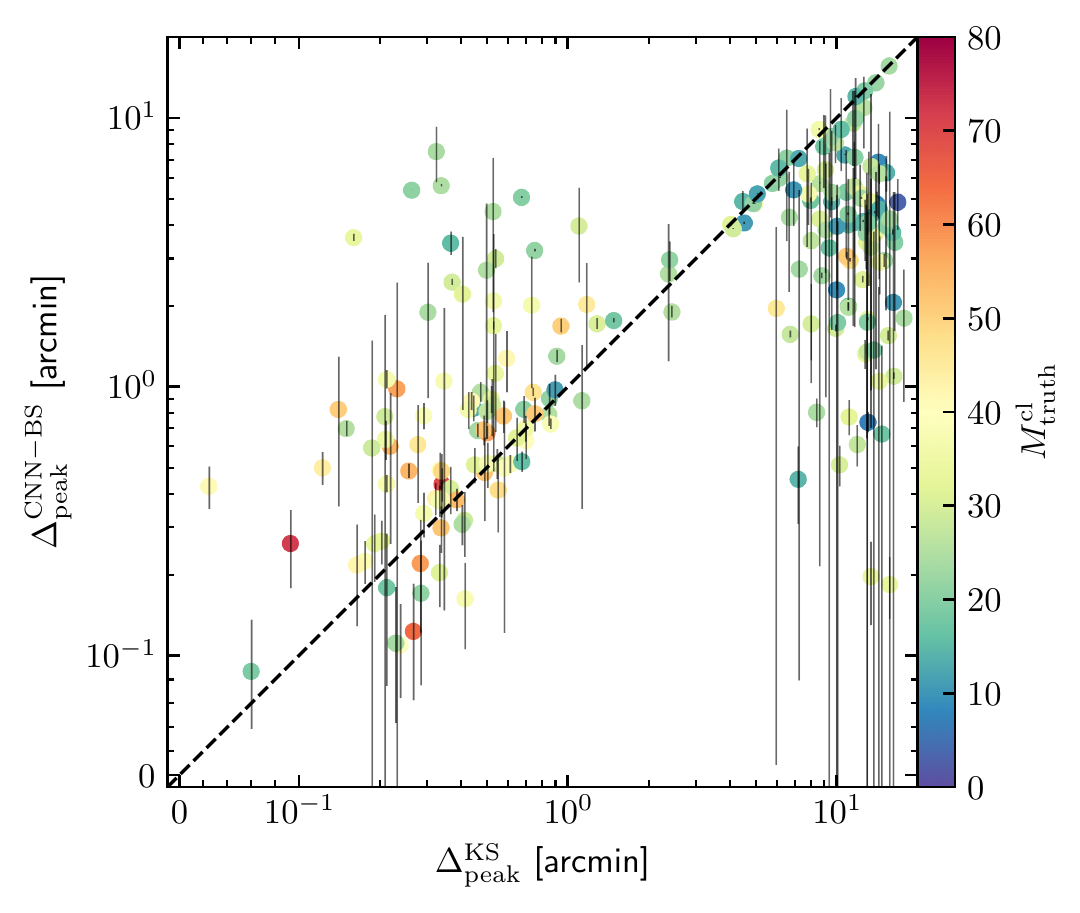}
\caption{CNN performance under the influence of bright-star masking. 
Left panel: joint probability measured from the convergence pixels within the masks.
Middle panel: comparison of projected cluster mass estimates with the truth.
Right panel: comparison of centroid deviations with the KS93 ones
performed without bright-star masking.
}
\label{fig:property_BS}
\end{figure*}

In order to examine the masking impact quantitatively, we measured the joint distribution, cluster mass comparison, and centroid distribution for the entire test sample as in \textsection\ref{sec:results_prob}, \textsection\ref{sec:results_mass}, and \textsection\ref{sec:results_peak}, respectively.
The joint $\kappa$ distribution displayed in the left panel of 
Figure~\ref{fig:property_BS} 
clearly indicates that the correlation with the truth in the high convergence regime is significantly weakened.
Compared with the non-masking case (right panel of Figure~\ref{fig:pixel}), the slope is reduced by a factor of
$1.5 \sim 2$ at $0.2\lesssim \kappa_{\rm truth} \lesssim 0.5$, which is consistent with our expectation from the visual inspection of the convergence map.

Since this weakened correlation in $\kappa$ is primarily due to the underestimation of the $\kappa$ values within the mask placed near the cluster center, we can expect that the
correlation in cluster mass also suffers in a similar fashion. 
The slope of the reconstructed mass to the truth becomes $0.554^{+0.257}_{-0.211}$ (see middle panel of Figure~\ref{fig:property_BS}), which is substantially smaller than the non-masking case $0.867^{+0.327}_{-0.296}$.

Finally, in terms of the centroid deviation, we find that the fraction of the catastrophic errors increases because the convergence values within the masked region are underestimated and this makes the largest convergence peak within the reconstructed field sometimes lie outside the masked area. The right panel of Figure~\ref{fig:property_BS} displays the comparison of the centroid deviation with the KS reconstruction result {\it performed without any masking}.
Even in the low deviation regime ($\Delta_{\rm peak}^{\rm KS}\lesssim10$ pixels), the CNN method
sometimes produces catastrophic errors. As mentioned earlier, this happens because of the in-mask underestimation. However, interestingly, in the regime where KS produces catastrophic errors ($\Delta_{\rm peak}^{\rm KS}\gtrsim30$ pixels), the CNN performance is sometimes significantly better. This may happen for the cases where the in-mask underestimation is less severe than the KS93 artifacts including 
noise amplification and inadequate $\kappa$-scale recovery (see the discussion in \textsection\ref{sec:results_visual}).

\section{Discussion} \label{sec:discussion}

\subsection{Why Does Our CNN algorithm Outperform the KS93 method?}

The comparison of our CNN mass reconstruction with the KS93 result has shown that the CNN performance is significantly better in several aspects (\textsection\ref{sec:results}). To name a few, the bias in the projected cluster mass estimation based on the convergence map is greatly reduced. And the fraction of catastrophic errors in the centroid measurement becomes much smaller especially in the low mass regime. Moreover, the convergence map is adaptively smoothed in such a way that a larger kernel is used in regions where the lensing S/N is lower, which leads to effective noise suppression in the cluster outskirts.
Here we present our discussion on the reason behind the outperformance.

The main cause for the improvement can be understood if we review some of the key issues in the conventional mass reconstruction (\textsection\ref{sec:conventional_MR}). The mass-sheet degeneracy is the most fundamental problem
because the shear $\gamma$ remains unchanged 
under the transformation of the convergence field:
$\kappa \rightarrow  \lambda \kappa + (1-\lambda)$.
This degeneracy can be lifted only by imposing some specific $\kappa$ value somewhere in the reconstruction field. One reasonable assumption is that the mean convergence is close to zero (although it should not be exactly zero) near the field boundary for a wide field mass reconstruction.
This allows us to determine the $\lambda$ value and thus break the degeneracy. Because our training data sets are drawn from cosmological simulation data, we believe that our CNN learns to utilize the information.

Another critical issue is the nonlinearity in the $g\rightarrow \kappa$ mapping. The fact that while the average ellipticity $\left < e \right >$ provides a reduced shear $g=\gamma/(1-\kappa)$, the convergence is a function of a shear $\gamma$ (equation~\ref{eqn:shear2kappa}) is ignored in the original KS93 formalism under the assumption that $g\sim \gamma$ (i.e., $\kappa \ll 1$) in the very weak gravitational lensing regime. Obviously, the condition $\kappa \ll 1$ is invalid in the typical cluster environment. Several suggestions are present in the literature to implement the nonlinearity.
For example, \citet{Seitz:1995dq} suggest an iterative procedure by updating $\gamma$ in equation~(\ref{eqn:shear2kappa}) with the information on $\kappa$ in the previous step. 
One drawback in this approach might be noise amplification through the iteration. Therefore, some authors propose maximum likelihood-based methods with some regularization constraints \citep[e.g.,][]{Seitz:1998mg,bradac2004,jee2007}. However, the fundamental limitation is that one needs $\kappa$ on an absolute scale in order to correctly address the nonlinearity $g=\gamma/(1-\kappa)$. That is, the nonlinearity problem cannot be addressed in isolation.
In CNN-based deep learning algorithms, these nonlinear issues are routinely addressed, and many applications turn out to be promising \citep[see, e.g.,][and references therein]{LIMK,MJU,RZGT}. In fact, the development of the CNN algorithm is motivated to address nonlinear problems such as denoising, image restoration, deconvolution, super-resolution, medical image reconstruction, holographic image reconstruction and so on. Therefore, it is not surprising to observe that combined with the mass-sheet degeneracy lifting capability, our CNN mass reconstruction significantly outperforms the original KS93 method.


\subsection{Test with Real Observations: Application to the El Gordo Cluster Data}

\begin{figure*}
\plotone{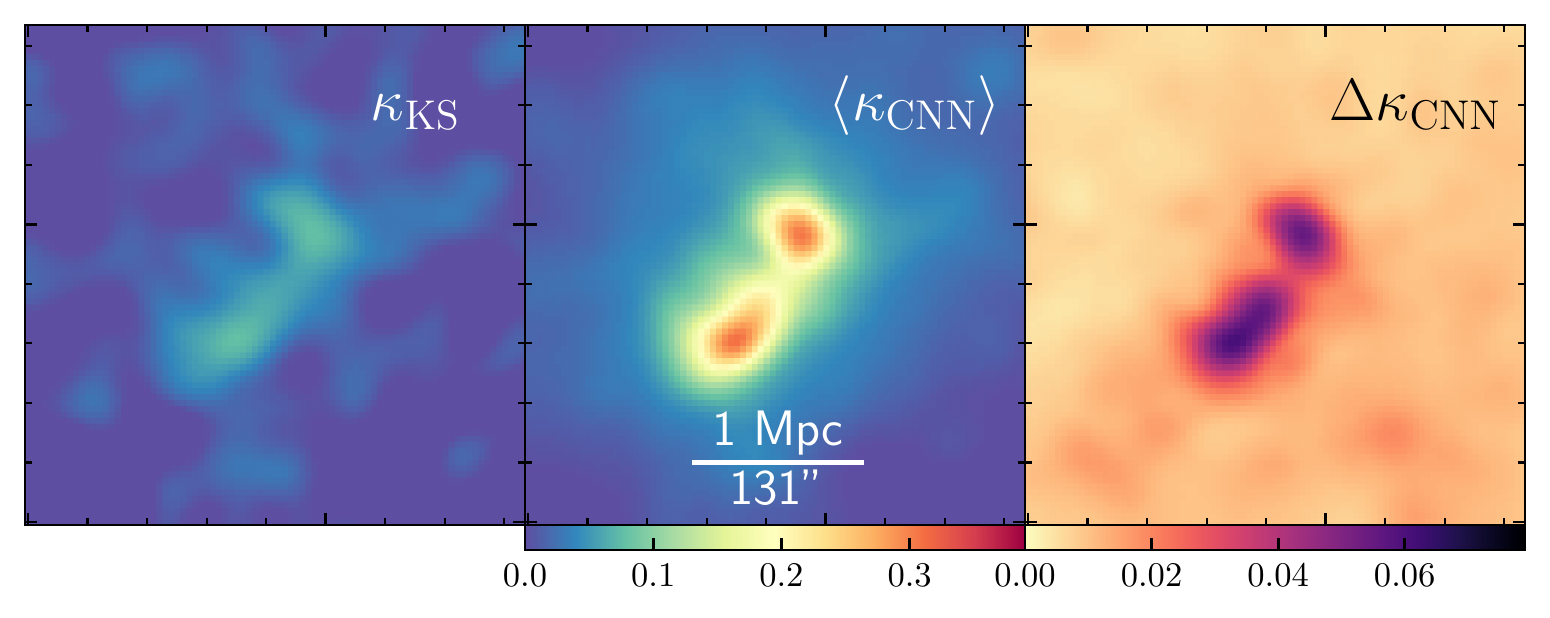}
\caption{Application of our CNN mass reconstruction to the
\emph{El Gordo} cluster. 
We use the {\it HST} WL catalog of \cite{Kim2021}. The left panel shows the mass reconstruction based on the KS93 algorithm. The CNN mass map in the middle panel is the average of the results from our 10 independent runs, which are also used to
estimate the standard deviation shown in the right panel.
Despite the differences between the training datasets and the El Gordo data, the CNN algorithm significantly outperforms the KS93 method in terms of the dynamical range representation, noise suppression, and substructure resolution.
}
\label{fig:visual_elGordo}
\end{figure*}

We have demonstrated that our CNN method can successfully reconstruct the projected mass maps from WL galaxy shears with the overall performance significantly better than that of the classical KS93 algorithm. Now one of the important questions is how well the current CNN method would work given real observational data, where a number of additional issues such as shear calibration errors, instrumental signatures, photometric redshift systematics, etc. are present.
With further development in deep learning and astronomical image generation tools, these issues may become tractable through end-to-end WL simulations in the future. Here we apply our CNN method to the 
{\it HST} WL data for the high-redshift merging cluster ``El Gordo" \citep{menanteau2014,jee2014}.
Within the current scope, we are interested in investigating the performance of our CNN method given the difference between the training data set and the real data in the following three aspects. First, our training was performed with the specific $32\arcmin \times 32 \arcmin$ field size whereas the {\it HST} field size ($\mytilde 9\arcmin\times9\arcmin$) of the El Gordo data is smaller by a few factors.
Second, the training is done by assuming that every source galaxy has an identical source redshift. Obviously, the source population in the El Gordo field comes from a wide range of redshifts and more importantly contains a significant fraction of non-background (contamination from cluster members and foreground objects) galaxies.
Third, the source density in the training data set is 25 per sq. arcmin, approximately a factor of four lower than the source density ($\mytilde100$ per sq. arcmin) in the {\it HST} observation of El Gordo.

Our HST catalog for El Gordo is provided by \cite{Kim2021}, who studied the cluster with a new wide-field HST imaging data set (PROP ID: 14153, PI. Hughes). Readers are referred to \cite{Kim2021} for details in the observation setup and reduction methods. In brief, the cluster was observed in four different programs (PROP IDs: 12477, 12755, 14096, and 14153). The entire field of view of the data with the addition of the last program (PROP ID: 14153) is $\mytilde119$ sq. arcmin, which covers the cluster beyond the virial radius $r_{200}\sim2$~Mpc. With the combination of all existing programs, the resulting average source density is $\mytilde95$ per sq. arcmin.

Figure~\ref{fig:visual_elGordo} displays the reconstructed mass map of El Gordo cluster from  our CNN model. The comparison with the KS93 version indicates that the advantages of the CNN method demonstrated in \textsection\ref{sec:results} with the simulated catalogs also manifest themselves here. First, the dynamical range of $\kappa$ is more realistic in the CNN version. El Gordo is one of the extremely massive clusters in the universe, and the projected convergence value should be $\kappa\gtrsim0.4$ in the central region based on the effective source redshift of $\mytilde1.2$ \citep{Kim2021} and the redshift of the cluster 0.87. 
The range of the convergence value in the CNN mass map is consistent with this expectation, although the relatively small field size does not allow us to lift the mass-sheet degeneracy completely. On the other hand, the peak convergence value in the KS93 case is too small.
Second, the KS93 inversion generates a number of spurious features in the outskirts whereas the CNN mass reconstruction efficiently suppresses these fluctuations. Currently, our multiwavelength data from X-ray to radio do not support the possibility that the features seen in the KS93 map might be real.
Third, the two mass peaks are better resolved in the CNN mass reconstruction. In the KS93 mass map, although one can see the presence of the two mass maps, there exists a ``bridge" connecting the
two mass peaks. Again, the existence of such a connecting substructure is not supported by our data.

\subsection{Null Test} \label{sec:null_test}
Although we take measures to prevent our CNN from learning that a cluster is always at the field center, our subsampling scheme for the
generation of the training dataset described in \ref{sec:cnn_architecture} still places the cluster
always within the central $\mytilde8.4$\% of the field ($\mytilde30$\% in each dimension).
Therefore, our CNN model constructed from this training dataset is expected to cause overestimation of the convergence in the central region.

\begin{figure}
\plotone{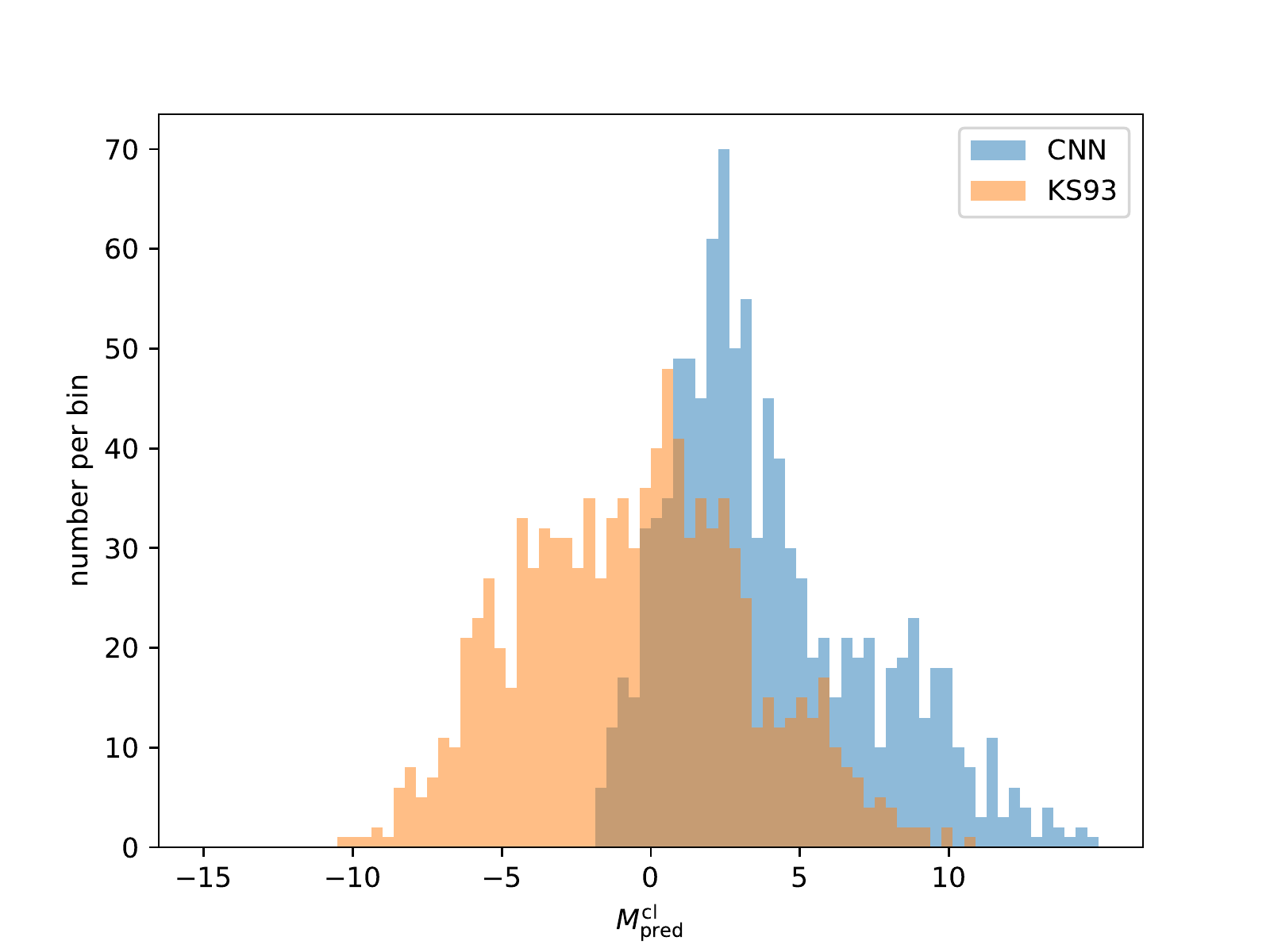}
\caption{Cluster mass null nest. We performed this null test by generating 1000 random galaxy catalogs for the null ($\kappa=0$) field and reconstructing the corresponding convergence fields with both the CNN and KS93 methods. We measured the projected mass from the $r=10$ pixel ($\mytilde0.72$~Mpc) circular aperture placed at the field center. While the KS93 result shows that the distribution is roughly symmetric around zero, our CNN mass clearly shows positive skewness. However, even the  cluster mass at the high-end tail ($M_{\rm pred}^{\rm cl}\sim 10$) corresponds to an insignificant cluster mass of $M_{200} \sim 7\times10^{13} M_{\sun}$, which is below the typical WL detection limit.}
\label{fig:convergence_bias}
\end{figure}

In order to quantify the bias, we performed a null test by generating 1000 random galaxy catalogs for the null ($\kappa=0$) field and reconstructing the corresponding convergence fields with both the CNN and KS93 methods.
We measured the projected mass from each convergence map using the $r=10$ pixel circular aperture placed at the field center. Figure~\ref{fig:convergence_bias} compares the distributions of the masses from CNN and KS93. The KS93 result shows that the distribution is roughly symmetric around zero. On the other hand, our CNN-based mass clearly shows positive skewness. This null test demonstrates that our CNN leads to overestimation of the convergence in the central region. Although not shown in  Figure~\ref{fig:convergence_bias}, we verify that the bias gradually decreases as we move the location of the aperture toward the edges.

We stress that this level of bias is insignificant in individual cluster mass estimation. For example, the projected mass at the high-end tail of the CNN distribution $M_{\rm pred}^{\rm cl}\sim 10$ corresponds to a cluster mass of $M_{200} \sim 7\times10^{13} M_{\sun}$, which is below the  detection limit in typical ground-based WL studies.
Nevertheless, we believe that future studies can reduce the bias substantially by improving the subsampling method and/or including blank fields in the training dataset.

\section{Conclusion} \label{sec:conclusion}

In this paper, we have introduced a new WL mass reconstruction method based on CNN algorithms.
Our CNN architecture consists of a series of 2D convolution and transposed-convolution layers
with implementation of skip-connections between the input and the output of the convolution-transposed-convolution layers via multiplication operation.
We generate training data sets using the ray-tracing data from cosmological simulations while the statistical properties of the source galaxies are designed to match those in our typical WL studies with Subaru/Suprime-Cam images.

Compared with the original KS93 inversion, our CNN method produces significantly improved results. The merits include enhancement in restoration of the dynamical range, agreement with the truth in both pixel-by-pixel and cluster mass comparisons, centroid determination, and noise suppression. In particular, it is remarkable that the slope of the recovered mass to the truth becomes consistent ($0.867^{+0.327}_{-0.296}$) with unity for the test sample. The slope is much lower ($0.484^{+0.218}_{-0.149}$) when we use the KS93 results instead. Also, we find that the centroid estimation based on the CNN result is much more stable in the low-mass regime.
We attribute these improvements to the efficient handling of the nonlinearity and degeneracy in our CNN algorithm, which however have been plaguing the traditional mass reconstruction methods.

The performance of our CNN algorithm somewhat degrades when a bright-star masking is fortuitously placed near the cluster center. Nevertheless, we find that the CNN reconstruction can still recover the cluster mass peak in most cases and the overall performance is still better than the KS93 results.

We tested our CNN model using the {\it HST} WL catalog of the El Gordo cluster. Despite the difference between our training data set and the real data in field size, source density, and redshift distribution, the CNN method clearly resolves the two mass clumps of the merging cluster in excellent agreement with the cluster member distribution while suppressing the noise fluctuation in the outskirts. 

Our study is the first implementation of WL mass reconstruction with CNN methods. Although further refinements in both algorithm and simulation are needed before we use the method for quantitative characterization of galaxy clusters, the result from this pilot study looks promising. One immediate application without the further improvements is shear-based galaxy cluster detection. Among the various selection methods, the shear-based galaxy cluster selection is unique in its ability to detect galaxy clusters with their projected masses. However, one of the most outstanding obstacles is the control of false positives due to noise fluctuation. As demonstrated throughout the paper, our CNN method suppresses this noise fluctuation efficiently while preserving the resolution in the high-density region, where the shear signal is high. In addition, since the projected $\kappa$ values are useful mass proxies, the CNN method can be used to provide the first classification of clusters according to their masses. Finally, the CNN-aided centroid determination and its comparison with other multiwavelength data can enhance our substructure identification and also reconstruction of merging scenarios in colliding clusters.

\acknowledgments
{The authors thank Inkyu Park, Cristiano Sabiu, David Parkinson, and Min-su Shin  for helpful discussions.
SEH, SP, and DB were (partly) supported by Basic Science Research Program through the National Research Foundation of Korea (NRF) funded by the Ministry of Education (2018\-R1\-A6\-A1\-A06\-024\-977).
SEH was also partly supported by the project \begin{CJK}{UTF8}{mj}우주거대구조를 이용한 암흑우주 연구\end{CJK} 
(``Understanding Dark Universe Using Large Scale Structure of the Universe''), funded by the Ministry of Science.
MJJ acknowledges support from the National Research Foundation of Korea under the program nos. 2017R1A2B2004644 and 2017R1A4A1015178.
DB was also supported by NRF Grant 2020\-R1\-A2\-B5\-B01\-001\-473. 
This work was also supported by the Korean Astronomy Machine Learning (KAML) working group.

This work is based on observations made with the NASA/ESA Hubble Space Telescope and operated by the Association of Universities for Research in Astronomy, Inc. under NASA contract NAS 5-2655.
Computational data were transferred through a high-speed network provided by the Korea Research Environment Open NETwork (KREONET).

We thank the Columbia Lensing group (\url{http://columbialensing.org}) for making their simulations available. The creation of these simulations is supported through grants NSF AST-1210877, NSF AST-140041, and NASA ATP-80NSSC18K1093.
We thank New Mexico State University (USA) and Instituto de Astrofisica de Andalucia CSIC (Spain) for hosting the Skies \& Universes site for cosmological simulation products.}

\software{Astropy \citep{astropy2013}, Keras \citep{keras2015}, Matplotlib \citep{matplotlib2007}, NumPy/SciPy \citep{scipy2020}, Pandas \citep{pandas2010}, SExtractor \citep{bertin1996}, \deleted{Theano \citep{theano2016},} Tensorflow \citep{tensorflow2015}}

\facility{HST (ACT)}


\appendix

\section{Performance Test with Other CNN Architectures}\label{sec:app_performance}

\begin{deluxetable}{lccc}[hbt]
\tablecaption{Same as Table~\ref{tab:performance}, including various CNN architectures.
The CNN model used in the main text is \textsf{FocalLoss}, and see  text for the definition of each architecture.
Those with the best performance for each parameter is marked as bold characters, while those with the worst performance for each parameter is underlined.
KS93 and \textsf{NoSkip} are not marked as underlined because they always show poorer performances than the other architectures.}\label{tab:performance_others}
\tablehead {
\colhead{Model} &
\colhead{$\mathcal{D}(\widetilde{\kappa}_{\rm pred},\widetilde{\kappa}_{\rm truth})$} &
\colhead{$M_{\rm pred}^{\rm cl}/M_{\rm truth}^{\rm cl}$} &
\colhead{$\Delta_{\rm peak}$}
}
\startdata
KS93 & $6.26 \pm 4.57$ & $0.484^{+0.218}_{-0.149}$ & 
$4\farcm27^{+8\farcm63}_{-4\farcm27}$ \\
\hline
\textsf{Fiducial} & $4.59 \pm 4.12$ & $0.762^{+0.305}_{-0.266}$ & 
$\mathbf{0\farcm55^{+4\farcm19}_{-0\farcm36}}$ \\
\textsf{4Channel} & $\underline{4.62 \pm 4.50}$ & $0.789^{+0.289}_{-0.263}$ & 
$0\farcm59^{+3\farcm66}_{-0\farcm37}$ \\
\textsf{19Filter} & $\mathbf{4.09 \pm 4.32}$ & $\underline{0.720^{+0.314}_{-0.238}}$ & 
$\underline{0\farcm73^{+1\farcm19}_{-0\farcm40}}$ \\
\textsf{29Filter} & $4.32 \pm 4.16$ & $0.754^{+0.305}_{-0.245}$ &
$0\farcm67^{+7\farcm31}_{-0\farcm44}$ \\
\textsf{FocalLoss} & $4.36 \pm 3.77$ & $\mathbf{0.867^{+0.327}_{-0.296}}$ &
$0\farcm60^{+4\farcm92}_{-0\farcm38}$ \\
\hline
\textsf{NoSkip} & $5.04 \pm 3.79$ & $-1.329^{+0.457}_{-0.842}$ & 
$0\farcm73^{+1\farcm19}_{-0\farcm40}$ \\
\enddata
\end{deluxetable}

\begin{deluxetable}{lccc}[hbt]
\tablecaption{Same to Table~\ref{tab:performance_others}, but for bright-star masking (see \textsection\ref{sec:results_star}).
\textsf{NoSkip} is not shown because of its poor performance.}
\label{tab:performance_others_bs}
\tablehead {
\colhead{Model} &
\colhead{$\mathcal{D}(\widetilde{\kappa}_{\rm pred},\widetilde{\kappa}_{\rm truth})$} &
\colhead{$M_{\rm pred}^{\rm cl}/M_{\rm truth}^{\rm cl}$} &
\colhead{$\Delta_{\rm peak}$}
}
\startdata
\textsf{Fiducial-BS} & $4.22 \pm 3.50$ & $0.448^{+0.217}_{-0.182}$ & 
$2\farcm60^{+4\farcm09}_{-2\farcm12}$ \\
\textsf{4Channel-BS} & $\underline{4.93 \pm 3.93}$ & $0.400^{+0.178}_{-0.135}$ & 
$4\farcm20^{+4\farcm48}_{-3\farcm67}$ \\
\textsf{19Filter-BS} & $4.91 \pm 3.90$ & $\underline{0.386^{+0.191}_{-0.149}}$ & 
$\underline{6\farcm05^{+4\farcm99}_{-5\farcm48}}$ \\
\textsf{29Filter-BS} & $4.29 \pm 3.55$ & $0.421^{+0.188}_{-0.193}$ & 
$2\farcm55^{+6\farcm09}_{-2\farcm11}$ \\
\textsf{FocalLoss-BS} & $\mathbf{3.66 \pm 3.22}$ & $\mathbf{0.554^{+0.257}_{-0.211}}$ & 
$\mathbf{1\farcm53^{+4\farcm21}_{-1\farcm14}}$ \\
\enddata
\end{deluxetable}

\begin{figure*}[hbt]
\centering
\plotone{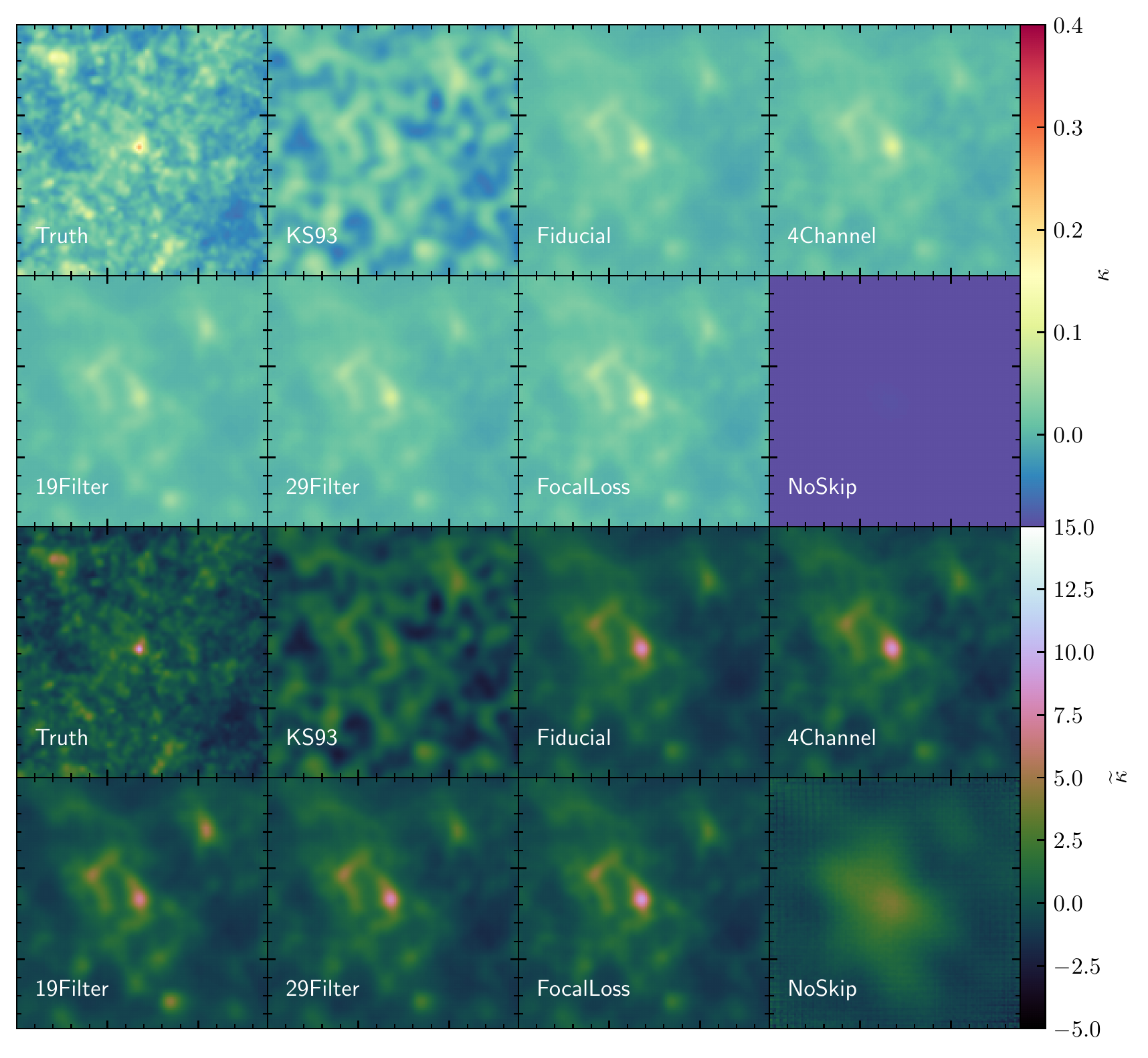}
\caption{Comparison of mass reconstruction results from different CNN architectures. See text for details of each variation. Most CNN variations show similar performances except for \textsf{NoSkip}, which suffers from significant resolution loss. According to our quantitative comparison based on the entire test sample, \textsf{FocalLoss} (used in the main text) produces the best results.  
}
\label{fig:visual_others}
\end{figure*}

We present performance tests of various CNN architectures using the diagnostics given in
\textsection\ref{sec:results}.
Hereafter the architecture presented in the main text of the current paper is referred to \textsf{FocalLoss}. It uses the modified mean-square-error loss function inspired by the focal loss \citep[][see equation~(\ref{eq:focal_loss})]{lin2017}.

In addition to \textsf{FocalLoss}, we test the following five variations:
\begin{description}
\item[\textsf{Fiducial}] same as \textsf{FocalLoss} except that the loss function is given by the standard mean square error (MSE):
\begin{equation}
\mathcal{L} = \sum_{\mathbf{x}} \left[ \kappa_{\rm pred}(\mathbf{x}) - \kappa_{\rm truth}(\mathbf{x}) \right]^2 \, .
\end{equation}
This MSE is also used for the rest of the variations.
\item[\textsf{4Channel}] same as \textsf{Fiducial} except that the smoothed number distribution of background galaxies is used as an additional channel of the input layer.
\item[\textsf{19Filter}] same as \textsf{Fiducial} except that the employed filter size is $19 \times 19$ (instead of $49 \times 49$) during the convolution and transposed convolution operations.
\item[\textsf{29Filter}] same as \textsf{Fiducial} except that the employed filter size is $29 \times 29$. 
\item[\textsf{NoSkip}] same as \textsf{Fiducial} except that it uses no skip-connection.
\end{description}

Figure~\ref{fig:visual_others} compares the mass reconstructions from these various CNN architectures. Judged by visual inspection, most CNN variations produce similar results. The exception is \textsf{NoSkip}, whose resolution is substantially compromised compared to the others.
When we examined the results from the individual runs, the \textsf{NoSkip} runs frequently produce null results, where the convergence map is flat ($\kappa(\mathbf{x}) \approx {\rm constant}$).
Also, the non-null results lack small-scale structures.
This comparison illustrates that the ResNet-like skip-connection between convolution and transposed-convolution layers plays a crucial role for recovering the details.
The \textsf{4Channel} result does not show any significant merit over \textsf{Fiducial}. This indicates that the additional information on the source  number distribution does not meaningfully contribute to the mass reconstruction quality.
In this example and also others, we find that the \textsf{19Filter} results tend to slightly overestimate the densities near the field edges compared to those produced with the larger ($29\times29$ or $49\times49$) filters, although the difference becomes insignificant when we compare the $29\times29$ vs $49\times49$ cases. This implies that there may exist a lower threshold in filter size in order to properly restore the dynamic range. 
Tables~\ref{tab:performance_others} and \ref{tab:performance_others_bs} summarize the comparisons among different CNN branches.

\bibliographystyle{apj}
\bibliography{article}


\end{document}